\documentclass[letterpaper,11pt]{article}
\pdfoutput=1

\usepackage{rotating}
\usepackage{jheppub}
\usepackage{slashed}
\usepackage{graphicx}
\usepackage{subfigure}
\usepackage{soul}
\usepackage{amsmath}
\usepackage{multirow}
\usepackage{tablefootnote}
\usepackage{hyperref}
\usepackage{xcolor}
\usepackage{epstopdf}
\usepackage{lscape}
\usepackage{url}
\usepackage{tikz}

\usepackage{ulem}

\usepackage{xcolor}
\usepackage{float}
\usepackage[utf8]{inputenc}

\newcommand{\beq}{\begin{equation}}
\newcommand{\eeq}{\end{equation}}
\newcommand{\bea}{\begin{eqnarray}}
\newcommand{\eea}{\end{eqnarray}}

\def\m1{M_1}
\def\m2{M_2}
\def\m3{M_3}

\def\ch10{\tilde \chi^0_1}

\def\to{\rightarrow}

\newcommand{\lsim}{\mathrel{\mathop{\kern 0pt \rlap
  {\raise.2ex\hbox{$<$}}}
  \lower.9ex\hbox{\kern-.190em $\sim$}}}
\newcommand{\gsim}{\mathrel{\mathop{\kern 0pt \rlap
  {\raise.2ex\hbox{$>$}}}
  \lower.9ex\hbox{\kern-.190em $\sim$}}}

\definecolor{pink}{RGB}{255,105,180}

\definecolor{green2}{rgb}{0,0.56,0.32}

\def\figureautorefname~#1\null{Fig.\,#1\null}
\def\tableautorefname~#1\null{Tab.\,#1\null}

\def\equationautorefname~#1\null{Eq.\,(#1)\null}

\title{The neutral scalars of type-II 2HDM+S under the LHC}

\author[a]{Cheng Li}
\author[a]{, Juxiang Li}
\author[b]{, Shufang Su}
\author[a]{and Wei Su}

\affiliation[a]{School of Science, Sun Yat-sen University, Gongchang Road 66, 518107 Shenzhen, China}
\affiliation[b]{Department of Physics, University of Arizona, Tucson, AZ 85721, U.S.A.}

\emailAdd{lich389@mail.sysu.edu.cn}
\emailAdd{lijx376@mail2.sysu.edu.cn}
\emailAdd{suwei26@mail.sysu.edu.cn}
\emailAdd{shufang@email.arizona.edu}

\abstract{The 2HDM+S is a singlet extension of the Two-Higgs-Doublet Model (2HDM), which offers rich collider phenomenology. In this paper, we parametrize the 2HDM+S with the Higgs masses and mixing angles, which provide a model-independent framework to study the collider signature.
Under five benchmark scenarios, we obtain the 95\% C.L. exclusion regions in the Type-II 2HDM+S parameter space by incorporating the SM-like 125~GeV Higgs precision measurements, beyond the Standard Model Higgs direct searches, $Z$-pole precision measurements and $B$-physics observables.  
We present the results in the Higgs boson masses vs $\tan\beta$, Higgs
boson masses vs mixing angles, $\tan\beta$ vs mixing angles and doublet Higgs
boson masses vs singlet Higgs boson mass parameter space.  We explore the complementarity between direct and indirect Higgs searches, as well as conventional Higgs search channels and exotic Higgs search channels. 
Compared to the 2HDM scenarios, we find that exotic channels such as $A/H \rightarrow Z h_S/ZA_S$  can probe large part of the parameter spaces, especially for moderate $1<\tan\beta<7$ region where the conventional channels in the 2HDM cannot contribute much. }

\keywords{Collider Phenomenology, Extended Higgs Sector.}

\begin{document}
\maketitle
\flushbottom
\section{Introduction}
The discovery of the Standard Model (SM)-like 125 GeV Higgs boson at the Large Hadron Collider (LHC)~\cite{ATLAS:2012yve,CMS:2012qbp} successfully confirmed prediction of the Standard Model. However, there are unsolved puzzles in particle physics, such as the existence of dark matter (DM), the non-zero neutrino mass, the baryon asymmetry of the universe, or the strong CP problem. Searches for new physics beyond the SM remain a frontier in particle physics research.  In various models to solve those puzzles, a non-SM Higgs sector is typically in need to modify the SM Higgs sector. The searches of extended Higgs sector beyond the Standard Model(BSM) becomes a crucial task at the LHC and future colliders. 

The Two Higgs Doublet Model with a singlet (2HDM+S) offers new sources for CP violation as well as possible DM candidates~\cite{Baum:2018zhf,Dutta:2023cig,Dutta:2025nmy}.  The Higgs sector of 2HDM+S includes two $\text{SU(2)}_L$ doublets and a complex singlet, sharing the same Higgs sector of the Next-to Minimal Supersymmetric Standard Model (NMSSM)~\cite{Ellwanger:2009dp} at low energy scale~\cite{Heinemeyer:2021msz,Biekotter:2021ovi}. Compared to the SM Higgs sector with only one physical Higgs $h$, the  Higgs spectrum of the 2HDM+S contains three CP-even BSM Higgs $h, H, h_S$, two CP-odd BSM Higgs $A, A_S$, and a pair of charged Higgses $H^\pm$ after the electroweak symmetry breaking. Additional symmetry is typically imposed to simplify the Higgs potential. For example, the singlet obey the $\mathbb{Z}_2$ symmetry of the 2HDM and additional $\mathbb{Z}^{'}_2$-odd symmetry leads to 2HDM + real Singlet Model (N2HDM), which can accommodate the possible 95~GeV excess at the LEP and the LHC~\cite{Heinemeyer:2021msz,Biekotter:2023oen}. In addition, the imposed $\mathbb{Z}_3$ symmetry in the Higgs potential would lead to the same Higgs structure as the NMSSM~\cite{Ellwanger:2009dp}.

The phenomenological studies of the 2HDM+S have only been done in some specific scenarios, while the more general cases of the 2HDM+S have not yet been explored in detail. In our study, we adopt the parametrization using the physical Higgs masses and mixing angles.  Such a model independent approach allows the mapping of our phenomenological results to models with a specific Higgs potential. To further simplify our analyses and capture the key features, we study five benchmark scenarios in which at most one mixing angle is set to be non-zero.  
In an early work~\cite{Li:2025zga}, we   examined the electroweak precision constraints on the 2HDM+S parameter space under five benchmark scenarios, as well as the complementarity between the electroweak and Higgs precision measurements.

There have been extensive searches for the BSM Higgses, such as conventional search channels of $H/A/H^\pm \to f\bar f^\prime$~\cite{DELPHI:hep-ex/0410017,ALEPH:hep-ex/0306033,ATLAS:1901.10917,ATLAS:1907.02749,CMS:1805.12191,CMS:1810.11822,ATLAS:2004.13605,CMS:1805.07504,CMS:2019pzc,CMS:1908.06463}, $VV'$~\cite{CMS:1912.01594,CMS:2109.06055,ATLAS:2009.14791,CMS:2109.08268} and $\gamma\gamma$~\cite{ATLAS:2102.13405,CMS:2405.09320,CMS:2405.18149}, as well as final states involving a SM-like Higgs $H \to Vh$~\cite{CMS:1807.02826,CMS:1903.00941,CMS:1910.11634,CMS:1911.03781,ATLAS:2004.01678,ATLAS:2011.05639,ATLAS:2110.13673,ATLAS:2207.00230,ATLAS:2023szc,CMS:2311.00130,CMS:2412.00570}, $hh$~\cite{CMS:2021roc,CMS:2024zfv,ATLAS:2022hwc,CMS:2024pjq,ATLAS:2020jgy,ATLAS:2022xzm,ATLAS:2020azv,CMS:2021yci,ATLAS:2023vdy,CMS:2024uru,ATLAS:2024ish,CMS:2022dwd,CMS:2024phk,ATLAS:2021ifb,ATLAS:2025hhd,CMS:2022xxa,CMS:2022fyt,ATLAS:2021hbr,ATLAS:2021ldb}. Once there is a mass hierarchy between the 2HDM Higgses, additional exotic decay modes, such as $H/A \to AZ/HZ$, $H/A\to H^\pm W^\mp$, $H \to AA, H^+H^-$ or $H^\pm \to HW/AW$ open up and quickly dominate the decay branching fractions. For the 2HDM+S, additional new channels open such as $A_S\to A h$, $h_S \to hh, AA, VV$, as well as channels with $h_S$ and $A_S$ in the final states: $H/A \to Z A_S/Zh_S$.
These additional decay channels relax the current collider limits on the heavy Higgses given the suppression of the decay branching fractions of those search channels, while offering new discovery channels at the same time. In our study, we analyze the 95\% C.L. exclusion region of the 2HDM+S parameter space under the five benchmark scenarios, taking into account the 125 GeV Higgs precision measurements, the electroweak precision measurements,  direct collider searches on the BSM Higgses, and flavor constraints.

 The rest of the paper is organized as follows. In \autoref{sec:thy}, we introduce the theoretical framework of 2HDM+S, the parametrization adopted in current study and the five benchmark cases. In \autoref{sec:cons}, we list the various experimental constraints we include in the current study. In \autoref{sec:results}, we present the results of the 95\% C.L. exclusion regions of the 2HDM+S parameter space under  five benchmark cases as well as the impact of the singlet mixing angles on the $\tan\beta$ vs $\cos(\beta-\alpha)$ parameter space. We conclude in \autoref{sec:conclu}.

%

\section{Theoretical framework}
\label{sec:thy}
The scalar fields  in the 2HDM with the singlet extension (2HDM+S)  contain
\begin{equation}
\Phi_1=\begin{pmatrix}
\chi_1^+\\ \frac{v_1+{\rho_1+i\eta_1}}{\sqrt{2}}
\end{pmatrix},\qquad\Phi_2=\begin{pmatrix}
\chi_2^+\\ \frac{v_2+{\rho_2+i\eta_2}}{\sqrt{2}}
\end{pmatrix},\qquad S= v_S + \rho_S + i\eta_S,
\end{equation}
where $\Phi_1$, $\Phi_2$ are the SU(2)$_L\times$U(1)$_Y$ doublets, and $S$ is the gauge singlet field. The neutral components of all scalar fields acquire the non-zero vacuum expectation value (vev), $v_1$, $v_2$ and $v_S$, and break the electroweak symmetry. The 2HDM+S Higgs potential is the combination of the 2HDM part and $V_S$ that contains interactions that involves the singlet field:
\begin{equation}
    V_\mathrm{2HDM+S} = V_\mathrm{2HDM}+V_\mathrm{S}.
\end{equation}
The 2HDM part is given by
\begin{equation}
    \begin{split}  V_\mathrm{2HDM}=&m_{11}^2\Phi_1^\dagger\Phi_1+m_{22}^2\Phi_2^\dagger\Phi_2-\left(m_{12}^2\Phi_1^\dagger\Phi_2+\mathrm{h.c.}\right)+\frac{\lambda_1}{2}(\Phi_1^\dagger\Phi_1)^2+\frac{\lambda_2}{2}(\Phi_2^\dagger\Phi_2)^2\\
		&+\lambda_3(\Phi_1^\dagger\Phi_1)(\Phi_2^\dagger\Phi_2)+\lambda_4(\Phi_1^\dagger\Phi_2)(\Phi_2^\dagger\Phi_1)\\&+(\Phi_1^\dagger\Phi_2)\left(\frac{\lambda_5}{2}(\Phi_1^\dagger\Phi_2) + {\lambda_6}(\Phi_1^\dagger\Phi_1)+ {\lambda_7}(\Phi_2^\dagger\Phi_2)\right)+\mathrm{h.c.}.
    \end{split}
\end{equation}
The singlet part of the Higgs potential has the most general formulation~\cite{Baum:2018zhf} as
\begin{equation}
	\begin{split}
		V_\mathrm{S}=&m_S^2 S^\dagger S + \frac{m_S'^2}{2}(S^2 + \mathrm{h.c.})+\Big(\frac{\mu_{S1}}{3!}S^3 + \frac{\mu_{S2}}{2}S(S^\dagger S) \\&+ S \Big[ \mu_{11} \Phi_1^\dagger\Phi_1 + \mu_{22} \Phi_2^\dagger\Phi_2 +\mu_{12}\Phi_1^\dagger\Phi_2 +\mu_{21}\Phi_2^\dagger\Phi_1 \Big] +\mathrm{h.c.}\Big)\\
        &+{S^\dagger S}\left(\lambda'_1\Phi_1^\dagger\Phi_1+\lambda'_2\Phi_2^\dagger\Phi_2 + {\lambda_3' \Phi_1^\dagger\Phi_2} +\mathrm{h.c.}\right)\\&+\Big[ {S^2}\left(\lambda'_4\Phi_1^\dagger\Phi_1+\lambda'_5\Phi_2^\dagger\Phi_2 +{\lambda'_6 \Phi_1^\dagger\Phi_2} + {\lambda'_7\Phi_2^\dagger\Phi_1} \right)+ \mathrm{h.c.}\Big]\\
		&  +\Big(\frac{\lambda_1''}{4!} S^4 + \frac{\lambda_2''}{3!} S^2 (S^\dagger S)+ \mathrm{h.c.}\Big)+{\frac{\lambda''_3}{4}(S^\dagger S)^2}.
	\end{split}
  \label{eq:hpot}
\end{equation}
Overall, this model has 28 free parameters.
However, imposing the additional symmetries on the scalar fields  reduces the number of parameters and lead to various structures of the Higgs potential. 
As in the 2HDM, in order to suppress the flavor changing neutral current (FCNC)\cite{Branco:2011iw}, the $\lambda_6$, $\lambda_7$, $\lambda_3'$, $\lambda_6'$, $\lambda_7'$ terms can be eliminated by imposing  the $\mathbb{Z}_2$ symmetry $\Phi_1\rightarrow -\Phi_1$, while the soft $\mathbb{Z}_2$ breaking terms $m_{12}$, $\mu_{12}$ and $\mu_{21}$ are still allowed.  Furthermore,  various symmetry structures are introduced in the literature to satisfy the theoretical constraints or achieve certain cosmological implications in the early universe. 
In Table~\ref{tab:symmetries}, we present a few symmetries considered in the literature, which  eliminate certain parameters  in the Higgs potential.  

\begin{table}[H]
\centering
\resizebox{\linewidth}{!}{  \begin{tabular}{cc|cc}
     Models& Symmetries& Remaining dimensionful terms& Remaining quartic terms\\
     \hline &&\\
               U(1)\cite{Biekotter:2021ovi}& $S\rightarrow e^{i\delta}S$& $m_{11},m_{22},m_{12},m_{S},m_S'$&$\lambda_1, \lambda_2,\lambda_3, \lambda_4, \lambda_5,\lambda_1',\lambda_2',\lambda_3''$\\ &&&\\
     $\mathbb{Z}_2'$\cite{Dutta:2023cig}& $S\rightarrow-S$& $m_{11},m_{22},m_{12},m_{S},m_S'$ & $\lambda_1, \lambda_2,\lambda_3, \lambda_4, \lambda_5,\lambda_1',\lambda_2',\lambda_1'',\lambda_3'',\lambda_4', \lambda_5'$\\&&&\\ 
     U(1)$_{PQ}$ \cite{Clarke:2015bea}& $S\rightarrow e^{\frac{i}{2}\alpha}S,\Phi_1\rightarrow e^{ic^2_\beta \alpha}\Phi_1$& $m_{11},m_{22},m_S$& $\lambda_1,\lambda_2,\lambda_3,\lambda_4,\lambda_1',\lambda_2',\lambda_3'',\lambda_7'$\\&$\Phi_2\rightarrow e^{-is^2_\beta \alpha}\Phi_2$&&\\ 
      & & &\\
          $\mathbb{Z}_3$\cite{Heinemeyer:2021msz}& {$S\rightarrow e^{-\frac{i2\pi}{3}}S,\Phi_2 \rightarrow e^{\frac{i2\pi}{3}}\Phi_2$}& $ m_{11},m_{22},m_{12},m_{S},\mu_{12},\mu_{S1}$&$\lambda_1, \lambda_2,\lambda_3, \lambda_4, \lambda_1',\lambda_2',\lambda_3''$
\end{tabular}  }
    \caption{Various symmetries considered in the literature on the 2HDM+S Higgs potential.}
    \label{tab:symmetries}
\end{table}

The collider phenomenology of the 2HDM+S, however,    has weak dependence on the specific symmetry of the Higgs potential. Since we only focus on the collider phenomenology and the mass eigenstates of the Higgs sector in this paper, we choose the Higgs potential with the $\mathbb{Z}_3$ \cite{Heinemeyer:2021msz} symmetry as a benchmark model in the following study.  

After electroweak symmetry breaking, the physical scalar mass spectrum contains three neutral CP-even scalars, two neutral CP-odd pseudoscalar and one pair of charged Higgs bosons, where the neutral Higgs bosons are an admixture of the two doublet fields and the singlet field. In order to reproduce the SM electroweak vacuum, the vevs of the doublet fields satisfy  $\sqrt{v_1^2 + v_2^2} = v\approx 246$~GeV, and we define $\tan\beta = \frac{v_2}{v_1}$. The mass matrices of neutral Higgs bosons are given in Eq.~\eqref{eq:masssmatrices} and Eq.~\eqref{eq:masspmatrices}.

For the CP-even states, the mixing matrix can be parametrized by three mixing angles
\begin{equation}
\begin{split}
  	R&= \begin{pmatrix}
       1& 0& 0\\
  	    0& c_{\alpha_{hS}}& s_{\alpha_{hS}}\\
       0& -s_{\alpha_{hS}}& c_{\alpha_{hS}}\\
  	\end{pmatrix}\begin{pmatrix}
  	    c_{\alpha_{HS}}& 0& s_{\alpha_{HS}}\\
       0& 1& 0\\
       -s_{\alpha_{HS}}& 0& c_{\alpha_{HS}}\\
       \end{pmatrix} \begin{pmatrix}
  	    c_{\alpha_{}}& s_{\alpha_{}}& 0\\
       -s_{\alpha_{}}& c_{\alpha_{}}& 0\\
       0& 0& 1
  	\end{pmatrix}\\
   &=\begin{pmatrix}
		c_{\alpha_{}}c_{\alpha_{HS}}& s_{\alpha_{}}c_{\alpha_{HS}}& s_{\alpha_{HS}}\\
		-s_{\alpha_{}}c_{\alpha_{hS}}-c_{\alpha_{}}s_{\alpha_{HS}}s_{\alpha_{hS}}& c_{\alpha_{}}c_{\alpha_{hS}}-s_{\alpha_{}}s_{\alpha_{HS}}s_{\alpha_{hS}}& c_{\alpha_{HS}}s_{\alpha_{hS}}\\
		s_{\alpha_{}}s_{\alpha_{hS}}-c_{\alpha_{}}s_{\alpha_{HS}}c_{\alpha_{hS}}& -s_{\alpha_{}}s_{\alpha_{HS}}c_{\alpha_{hS}}-c_{\alpha_{}}{{s_{\alpha_{hS}}}}& c_{\alpha_{HS}}c_{\alpha_{hS}}
	\end{pmatrix},
 \label{eq:roteven}
\end{split}
\end{equation} 

where
\begin{equation}
R    \begin{pmatrix}
        \rho_1\\ \rho_2\\ \rho_S
    \end{pmatrix} = \begin{pmatrix}
        H\\ h\\ h_S
    \end{pmatrix},\qquad R M_S^2 R^T =  \operatorname{diag}\{m_H^2, m_{h}^2, m_{h_S}^2\}.
    \label{eq:roteven2}
    \end{equation}
The three physical CP-even Higgs bosons are the SM-like doublet Higgs boson $h$, the doublet-like Higgs boson $H$, and the singlet-like Higgs boson $h_S$.   The three mixing angles are the angle $\alpha$ for the mixing between two 2HDM CP-even fields,  $\alpha_{hS}$ for the mixing between the light 2HDM Higgs and the singlet field $\rho_S$, and $\alpha_{HS}$ for the mixing between the heavy 2HDM Higgs and singlet field $\rho_S$. 
For the CP-odd states, we have 
\begin{equation}
\begin{pmatrix}
        G^0\\ A \\ A_S
    \end{pmatrix}=       \begin{pmatrix}
        1& ~0& ~0\\
        0& \multicolumn{2}{c}{\multirow{2}{*}{$R^A$}}& \\
        0& &  \\
    \end{pmatrix}
    \begin{pmatrix}
                c_{\beta}& s_{\beta}& 0\\
        -s_{\beta}& c_{\beta}& 0\\
        0 & 0& 1
    \end{pmatrix}\begin{pmatrix}
        \eta_1 \\ \eta_2 \\ \eta_S
    \end{pmatrix},\qquad
    R^A = \begin{pmatrix}
        c_{\alpha_{AS}}& s_{\alpha_{AS}}\\
        -s_{\alpha_{AS}}& c_{\alpha_{AS}}
    \end{pmatrix},
\end{equation}
where $G^0$ is the neutral Goldstone boson eaten by $Z$, and the angle $\alpha_{AS}$ is the mixing angle between 2HDM doublet pseudoscalar $A_{2HDM}$ and the singlet pseudoscalar $\eta_S$. The charged Higgs sector is the same as the one in the 2HDM with one pair of Goldstone bosons $G^\pm$ eaten by $W^\pm$ and one pair of charged Higgs bosons $H^\pm$: 
\begin{equation}
    \begin{pmatrix}
        G^+\\ H^+
    \end{pmatrix} = \begin{pmatrix}
        c_\beta&
s_\beta\\
-s_\beta&   c_\beta\end{pmatrix}\begin{pmatrix}
    \chi^+_1\\ \chi^+_2
\end{pmatrix}.
\end{equation}

The free parameters of 2HDM+S with $\mathbb{Z}_3$ symmetry in the interaction basis can also be written in terms of the mass eigenvalues and the mixing angles,  as presented in Table~\ref{tab:inpars}.
\begin{table}[h]
    \centering
    \resizebox{\linewidth}{!}{\begin{tabular}{l|ccccccccccccc}
        Basis &  \multicolumn{12}{c}{Input parameters} \\
        \hline
         Interaction eigenstate&  $\tan\beta$& $v_S$&$m_{12}$&$\lambda_1$&$\lambda_2$&$\lambda_3$&$\lambda_4$& $\lambda_5$&$\lambda_1'$& $\lambda_2'$& $\lambda_3''$& $\mu_{S1}$& $\mu_{12}$ \\
         Mass eigenstate& $\tan\beta$&$v_S$&$m_{\phi}$&$m_H$&$m_h$&$\alpha$&$m_{H^\pm}$& $m_A$& $\alpha_{HS}$& $\alpha_{hS}$& $m_{h_S}$& $m_{A_S}$& $\alpha_{AS}$
    \end{tabular}}
    \caption{The  free parameters of 2HDM+S with $\mathbb{Z}_3$ symmetry in the interaction eigenstate basis and the mass eigenstate basis.}
    \label{tab:inpars}
\end{table}
 
The Higgs to gauge bosons couplings can be obtained by
\begin{equation}
    c_{h_i VV}=c_{h_iWW}=c_{h_iZZ} = R_{i1} c_\beta + R_{i2}s_\beta,
\end{equation}
where $c_{h_iXX}$ is the reduced coupling, which is the coupling of Higgs bosons to two SM particles rescaled by the SM values: 
\begin{equation}
    c_{h_i XX} = \frac{g_{h_iXX}}{g_{h_\mathrm{SM}XX}}.
\end{equation}
By applying the mixing matrix of Eq.~\eqref{eq:roteven}, the CP-even Higgs to gauge bosons couplings are given by Table~\ref{tab:hvvcoups}.
\begin{table}[H]
    \centering
    \begin{tabular}{lc}
 & $c_{h_i VV}$ \\
\hline
$H$& $c_{\beta-\alpha}c_{\alpha_{HS}} $\\
$h$& $s_{\beta-{\alpha}}c_{\alpha_{hS}} - c_{\beta-{\alpha}}s_{\alpha_{HS}}s_{\alpha_{hS}} $\\
$h_S$& $-s_{\beta-\alpha}s_{\alpha_{hS}} - c_{\beta-\alpha}s_{\alpha_{HS}}c_{\alpha_{hS}}$\\
    \end{tabular}
    \caption{The Higgs bosons to gauge bosons couplings in  the 2HDM+S.}
    \label{tab:hvvcoups}
\end{table}

In the mass eigenstate basis, the reduced Higgs to fermions couplings are shown in Table~\ref{tab:yuks} for four different types of Yukawa structure. 
\begin{table}[h]
    \centering
    \begin{tabular}{c|cccc}
    & type-I&   type-II&    type-X&    type-Y\\
    \hline
         $c_{h_i uu}$& $R_{i2}/s_\beta$& $R_{i2}/s_\beta$& $R_{i2}/s_\beta$& $R_{i2}/s_\beta$\\
         $c_{h_i dd}$& $R_{i2}/s_\beta$& $R_{i1}/c_\beta$& $R_{i2}/s_\beta$& $R_{i1}/c_\beta$\\
         $c_{h_i \ell\ell}$& $R_{i2}/s_\beta$& $R_{i1}/c_\beta$& $R_{i1}/c_\beta$& $R_{i2}/s_\beta$\\
    \end{tabular}
    \caption{Higgs to fermions couplings for different Yukawa structure types.}
    \label{tab:yuks}
\end{table}

The general phenomenology of 2HDM+S could be quite complicated given the multiple mixing angles.  In our analyses below, we focus on five benchmark cases with at most one mixing angle is set to be non-zero \cite{Li:2025zga}:
\begin{table}[h]
    \centering
    \begin{tabular}{llll}
    \hline
    \multicolumn{2}{l}{Benchmark}  &   Fixed mixing angles&    Varying mixing angle\\
    \hline
         Case-0& & $c_{\beta-\alpha}=\alpha_{HS}=\alpha_{hS}=\alpha_{AS}=0$&~~~~ ---, alignment limit of 2HDM\\
         Case-I& & $\alpha_{HS}=\alpha_{hS}=\alpha_{AS}=0$& ~~~~$c_{\beta-\alpha}$, 2HDM limit\\
         Case-II& & $c_{\beta-\alpha}=\alpha_{HS}=\alpha_{AS}=0$& ~~~~$\alpha_{hS}$: $h-h_S$ mixing\\
         Case-III& & $c_{\beta-\alpha}=\alpha_{hS}=\alpha_{AS}=0$& ~~~~$\alpha_{HS}$: $H-h_S$ mixing\\
         Case-IV& & $c_{\beta-\alpha}=\alpha_{hS}=\alpha_{HS}=0$& ~~~~$\alpha_{AS}$: $A-A_S$ mixing\\
         \hline
    \end{tabular}
    \caption{Five benchmark cases for the mixing angles.}
    \label{tab:limits}
\end{table}
\begin{description}
    \item[Case-0] $c_{\beta-\alpha}=\alpha_{HS}=\alpha_{hS}=\alpha_{AS}=0$ is the 2HDM alignment limit, where the singlet components are decoupled and 125~GeV Higgs $h$ is the same as the SM Higgs. 
    \item[Case-I] $\alpha_{HS}=\alpha_{hS}=\alpha_{AS}=0$ is the 2HDM limit, when the singlet components are completely decoupled. 
    \item[Case-II] $\alpha_{hS} \neq 0$ and $c_{\beta-\alpha}=\alpha_{HS}=\alpha_{AS}=0$ represents the case when the 125 GeV $h$ mixes with the singlet Higgs $h_S$, and the BSM doublet Higgses $H/A/H^\pm$ are decoupled with $h_S$.
    \item[Case-III] $\alpha_{HS} \neq 0$ and $c_{\beta-\alpha}=\alpha_{hS}=\alpha_{AS}=0$ represent the case when the BSM $H$ mixes with the singlet Higgs $h_S$, while the 125~GeV Higgs $h$ is completely SM-like.
    \item[Case-IV] $\alpha_{AS} \neq 0$ and $c_{\beta-\alpha}=\alpha_{HS}=\alpha_{hS}=0$ represent the case when $A$ mixes with the singlet pseudoscalar $A_S$, where the CP-even sector is the same as the alignment limit of the 2HDM, with a decoupled singlet scalar $h_S$.
\end{description}

In this study, we focus on the type-II Yukawa coupling structure, which corresponds to the Yukawa structure of the supersymmetric models, e.g. NMSSM and MSSM.
The couplings of the 2HDM+S Higgses to SM particles, including fermions and gauge bosons couplings, are shown in Table~\ref{tab:hff_hvcoups} in Appendix~\ref{sec:hcpls} for both the general form and the specific couplings in five benchmark cases.
Furthermore, the general formulae of trilinear Higgs couplings, including $g_{h_ih_jh_k}$, $g_{h_i a_j a_k}$, $g_{h_i H^+H^-}$, are presented in Eqs.~\eqref{eq:hhh},~\eqref{eq:haa} and \eqref{eq:hhphm}, respectively. These couplings under the five benchmark cases are  presented in Tables~\ref{tab:hhh_01234} and~\ref{tab:haamp}.  
\section{Collider constraints on various scenarios}
\label{sec:cons}

The parameter space of the 2HDM+S are constrained by theoretical considerations and experimental measurements. The theoretical constraints, including the tree-level perturbative unitarity \cite{Horejsi:2005da}, potential bounded from below \cite{Klimenko:1984qx} and the vacuum stability \cite{Branchina:2018qlf}, strongly depend on the explicit symmetry structure of the Higgs potential. 
For the 2HDM+S with $\mathbb{Z}_2$ and $\mathbb{Z}_3$ symmetries, the conditions that satisfy the theoretical considerations are obtained in \cite{Heinemeyer:2021msz,Dutta:2023cig}. Imposing the theoretical considerations and successful electroweak symmetry breaking might  restrict the mixing angles to a smaller range. These theoretical constraints, however, depend strongly on the choice of $m_{12}$ and $v_S$, which are less relevant for the collider phenomenology. In the following study, we always set $ m_H = m_A=m_{H^\pm}=m_\phi$ ($m_\phi$ as defined \autoref{eq:mphi}) to avoid most of the theoretical constraints. For specific 2HDM+S model with certain symmetry imposed on the Higgs potential, we have ran parameter scan with theoretical constraints imposed. We found that a wide range of masses as well as mixing angles can be realized.  Therefore, in our study below, we focus on the experimental constraints as follows.




\begin{description}
    \item [The 125~GeV Higgs precision measurements] A SM-like 125~GeV Higgs boson has been observed at the LHC in various channels including $WW$~\cite{CMS:2022uhn,ATLAS:2025hki}, $ZZ$~\cite{ATLAS:2020rej,CMS:2021ugl}, $bb$~\cite{ATLAS:2024yzu,CMS:2024ddc}, $\gamma\gamma$~\cite{CMS:2021kom,ATLAS:2022tnm} and $\tau\tau$~\cite{ATLAS:2022akr,CMS:2022kdi}, with the measurements of the signal strength of specific channel $i$ as
    \begin{equation}
        \mu_i  = \frac{ (\sigma\cdot\mathrm{BR})_i}{(\sigma^\mathrm{SM}\cdot\mathrm{BR}^\mathrm{SM})_i}.
    \end{equation}

   We use the public code \texttt{HiggsSignals~2}~\cite{Bechtle:2020uwn}, which is embedded in \texttt{HiggsTools}~\cite{Bahl:2022igd}. The \texttt{HiggsSignals~2} calculates the following likelihood function
    \begin{equation}
        \chi^2_\mathtt{HS} = \sum_{ij}(\hat{\mu_i}-\mu_i){C^{-1}_\mu}_{ij}(\hat{\mu_j}-\mu_j),
    \end{equation}
    where $\hat{\mu_i}$ is the prediction of the signal strength in a particular model and the ${C^{-1}_\mu}_{ij}$ includes the correlation coefficients between different channels.    We  determine the 95\% C.L. (Confidence Level) allowed region of the  BSM prediction for the signal strength of the 125~GeV Higgs as 
    \begin{equation}
        \Delta\chi^2 = \chi^2_\mathtt{HS} - {\chi^2_\mathtt{HS}}(\mathrm{SM}) < 5.99.
    \end{equation}
  While the mixing angle of $\cos(\beta-\alpha)$ in the Type-II 2HDM has been constrained tightly by the current 125 GeV Higgs coupling measurements~\cite{Han:2020lta,Bahl:2022igd}, its range in the 2HDM+S could be shifted when mixing with the singlet Higgs is non-zero~\cite{Li:2025zga}.  

    \item [BSM Higgs direct searches] Given the negative search results for the BSM particles, the additional Higgs bosons in the 2HDM+S should not exceed the 95\% C.L.   bounds of the experimental scalar particle searches. We use \texttt{HiggsBounds~5}~\cite{Bechtle:2020pkv,Bahl:2021yhk} embedded in \texttt{HiggsTools}~\cite{Bahl:2022igd} to calculate the cross-section times branching fractions observed for explicit BSM Higgs search channels
    \begin{equation}
        r_\mathrm{obs}=\frac{(\sigma\cdot\mathrm{BR})_\mathrm{pred}}{(\sigma\cdot\mathrm{BR})_\mathrm{95\% C.L. limit}}.
    \end{equation}
    The BSM Higgs in the 2HDM+S would be allowed by the experimental searches at 95\% C.L.   when the ratio $r_\mathrm{obs}<1$.
    \item [Electroweak precision observables] The oblique parameters $S$, $T$ and $U$ obtained from the electroweak precision observables are sensitive to the self energies of the $W^\pm$ and $Z$ bosons. The contributions from the BSM Higgses are highly constrained by the current electroweak precision measurements~\cite{stupdg:2024}. We obtain the $S$, $T$ and $U$ parameters in the 2HDM+S by adapting the calculation of $STU$ for multi-Higgs-doublets models~\cite{Grimus:2008nb}. The implication of the electroweak precision constraints on the 2HDM+S has been studied in our earlier work~\cite{Li:2025zga}, which is included in the current study.
    \item [Flavor  observables] Additional neutral and charged Higgs bosons in the 2HDM+S could contribute to  flavor observables such as the branching ratio of $B\rightarrow X_s \gamma$, $B$ meson oscillation parameter $\Delta m_{B_s}$, and  $B_s\rightarrow\mu^+\mu^-$ etc. 
    The 2HDM parameter space of $m_{H^\pm}\lesssim 750$~GeV or $\tan\beta \lesssim 1$ is excluded for the Type-II Yukawa coupling structure~\cite{Li:2024kpd}.  Such flavor constraints could also be relaxed in the 2HDM+S scenario.
\end{description}

%
\section{Phenomenological analyses for the parameter space}
\label{sec:results}
For our study, we used \texttt{SARAH}~\cite{Staub:2013tta} to generate the 2HDM+S model file, which is used by  \texttt{SPheno}~\cite{Porod:2011nf} to calculate all the Higgs bosons decay branching fractions.
The Higgs production cross sections are calculated using the \texttt{HiggsPredictions} in \texttt{HiggsTools}~\cite{Bahl:2022igd}.   We compared with the current electroweak and Higgs precision measurements, as well as LHC direct search limits on the BSM Higgses and flavor constraints to obtain the 95\% C.L. exclusion regions on the 2HDM+S parameter spaces.

We performed our analyses in the five benchmark cases listed  in Table~\ref{tab:limits}.  We present the parameter space of $\tan\beta$ vs Higgs boson masses in the section~\ref{sec:tbvm},  mixing angles vs Higgs boson masses in section~\ref{sec:avm}, $\tan\beta$ vs mixing angles in section~\ref{sec:tbva},  and the doublet Higgs boson masses vs singlet Higgs boson mass in section~\ref{sec:mvm}.

\subsection{$\tan\beta$ - $m_{h_i}$}
\label{sec:tbvm}
\begin{figure}[h]
    \centering
    \includegraphics[width=.5\linewidth]{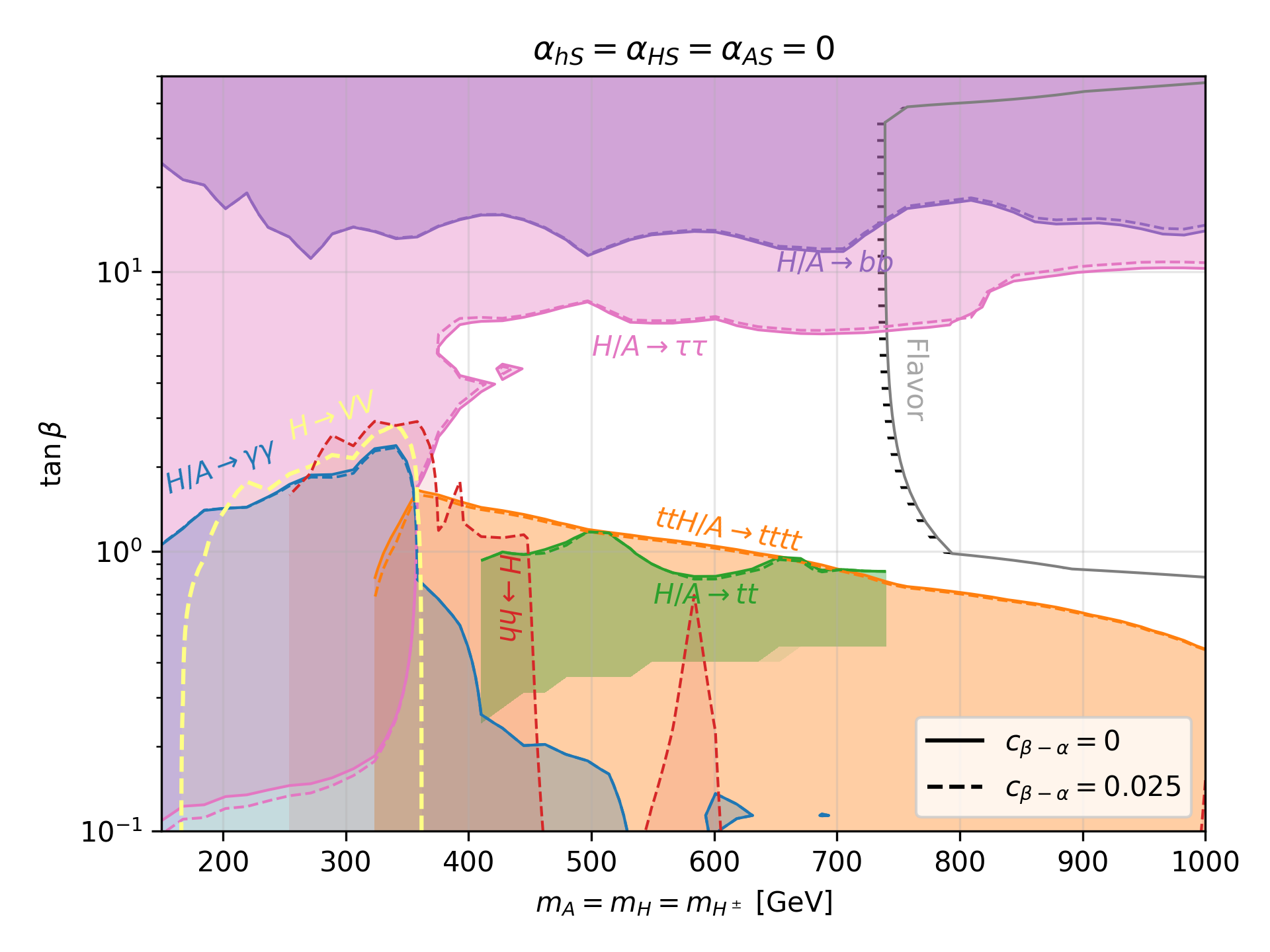}\includegraphics[width=.5\linewidth]{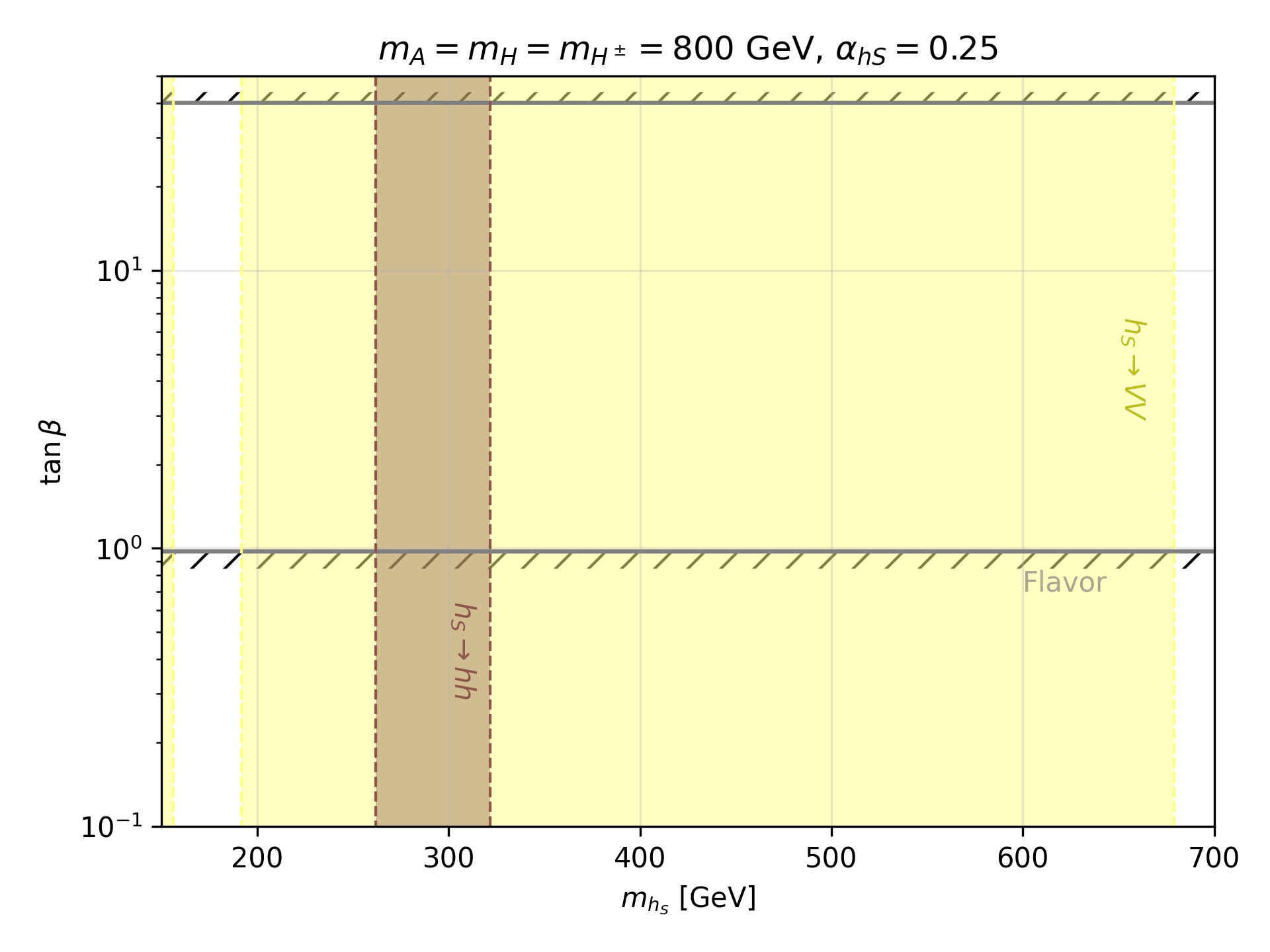}
    \includegraphics[width=0.5\linewidth]{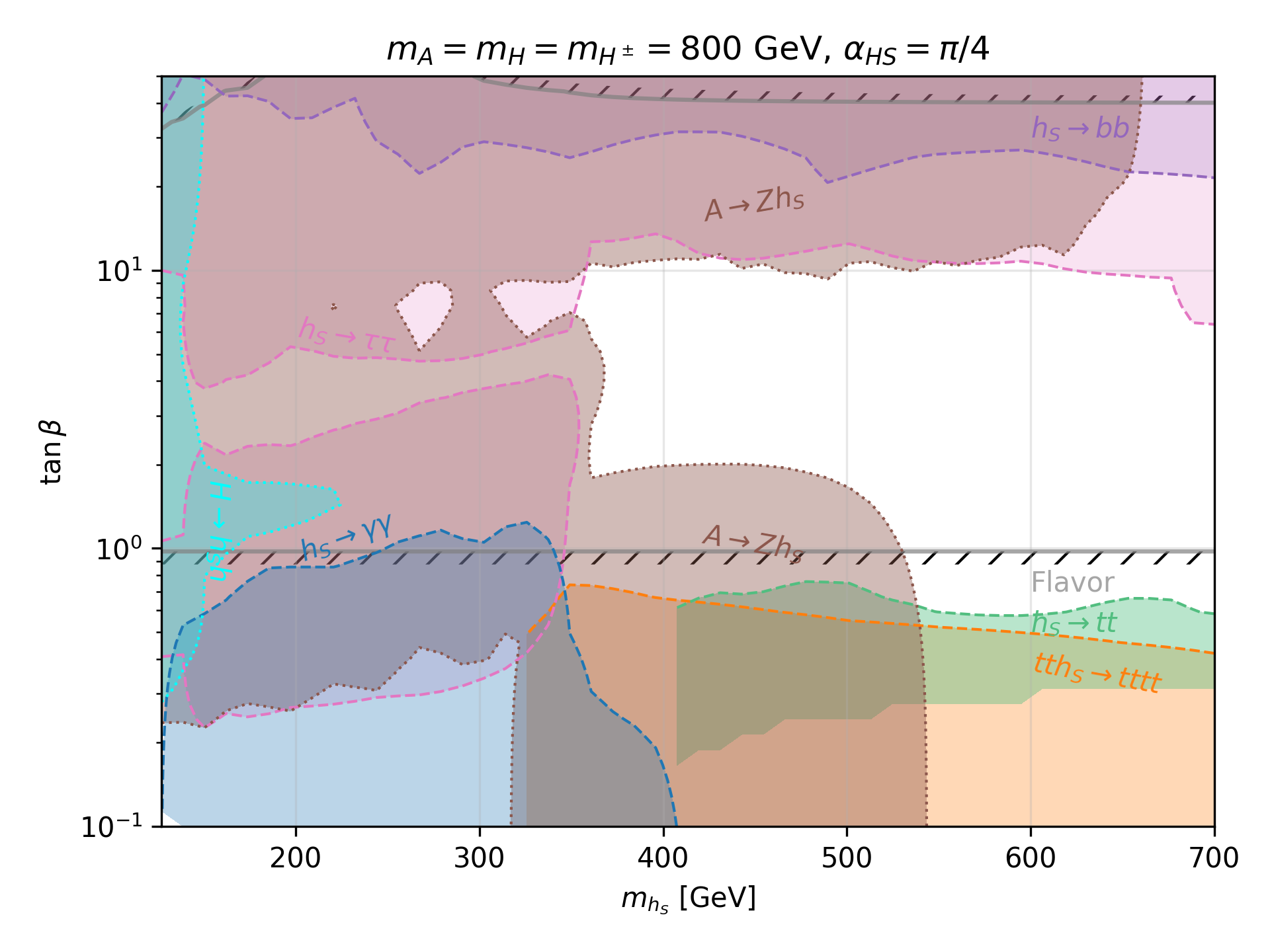}\includegraphics[width=0.5\linewidth]{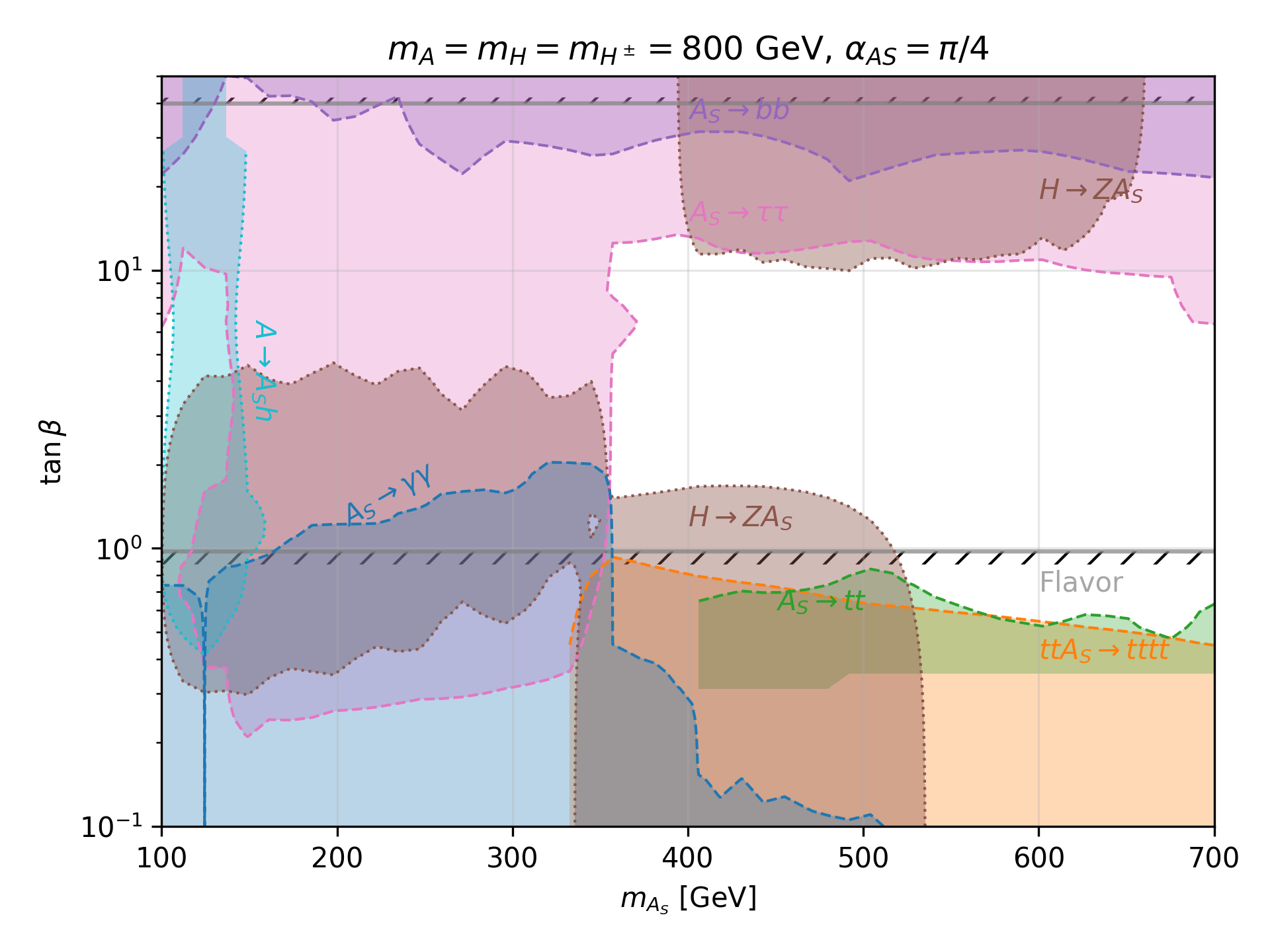}
    \caption{The 95\% C.L. exclusion regions of the  $\tan\beta$ vs different Higgs masses parameter space in five benchmark cases, with $v_S=v$. The upper left panel is $\tan\beta$ vs $m_{A/H/H^\pm}$ in  case-0 (solid lines) and Case-I (dashed lines) with $m_{h_S}=m_{A_S}=1.5$~TeV. The upper right panel is $\tan\beta$ vs $m_{h_S}$ in  case-II with $\alpha_{hS}=0.25$. The lower left panel is $\tan\beta$ vs $m_{h_S}$ in  case-III with $\alpha_{HS}=\frac{\pi}{4}$. The lower right panel is $\tan\beta$ vs $m_{A_S}$ in  case-IV with $\alpha_{AS}=\frac{\pi}{4}$. For the Case-II $-$ IV, $m_{A/H/H^\pm}$ is set to be 800~GeV.}
    \label{fig:tb_ma}
\end{figure}

Fig.\ref{fig:tb_ma} shows 95\% C.L. exclusion regions in $\tan\beta$ vs BSM Higgs bosons masses parameter space. The top-left panel shows  Case-0 (solid curves) and Case-I (dashed curve with $c_{\beta-\alpha}=0.025$) with varying degenerated BSM Higgs masses $m_H=m_A = m_{H^\pm}$.
The results for the Case-0  are the same as in Ref.~\cite{Kling:2020hmi}. $\tau\tau$ channel rules out the large $\tan\beta$ region, as well as mass region below the $t\bar{t}$ threshold, while the $tttt$ continuum search is sensitive for the low $\tan\beta$ region.
For $c_{\beta-\alpha}\neq 0$, the $H\rightarrow VV$ channel and $H\rightarrow hh$ channel open up, which are sensitive at the low $m_H$ and low $\tan\beta$ region, as shown by the yellow and red dashed lines.  Given the tight constraint on $c_{\beta-\alpha}$ due to precision measurements of the SM-like Higgs, a small value of $c_{\beta-\alpha}=0.025$ does not have a significant effect on the fermionic channels, as shown by the almost overlap solid and dashed lines for those channels.  Regarding the flavor constraints, the $B\to X_S\gamma $ would exclude the region of $m_{H^\pm}\lesssim750$~GeV while $B_s\to\mu^+\mu^-$ would rule out the region of $\tan\beta <1$ and $\tan\beta>40$. The flavor constraints, however, dominantly depend on the $m_{H^\pm}$ and $\tan\beta$, where the mixing angles in the neutral sector do not play a role. For the study of other benchmark cases, we fix the charged Higgs mass to be above 800 GeV.  The flavor observables would constrain the $\tan\beta$ within $[1,~40]$ consequently.

The top-right panel shows  Case-II with varying $m_{h_S}$, for  $m_H=m_A = m_{H^\pm}=800$ GeV and $\alpha_{hS}=0.25$. The value of $\alpha_{hS}$ is chosen so that it is not excluded by the current SM-like Higgs precision measurements.   The limits of $h_S \rightarrow VV$ and $h_S\rightarrow hh$ are independent of $\tan\beta$, since ${h_S}$ mixes with the SM-like $h$ and the properties of $h$ are independent of $\tan\beta $ at the alignment limit.  Constraints from the fermionic decay channels of $h_S$ are relatively weak. We did not show the limits of the doublet Higgs $A$ and $H$ channels ($bb$, $\tau\tau$, $tttt$) since the corresponding limits highly depend on the benchmark value of $m_{A/H}$, as shown in the upper left panel, while independent of $m_{h_S}$.

\begin{figure}
    \centering
    \includegraphics[width=0.5\linewidth]{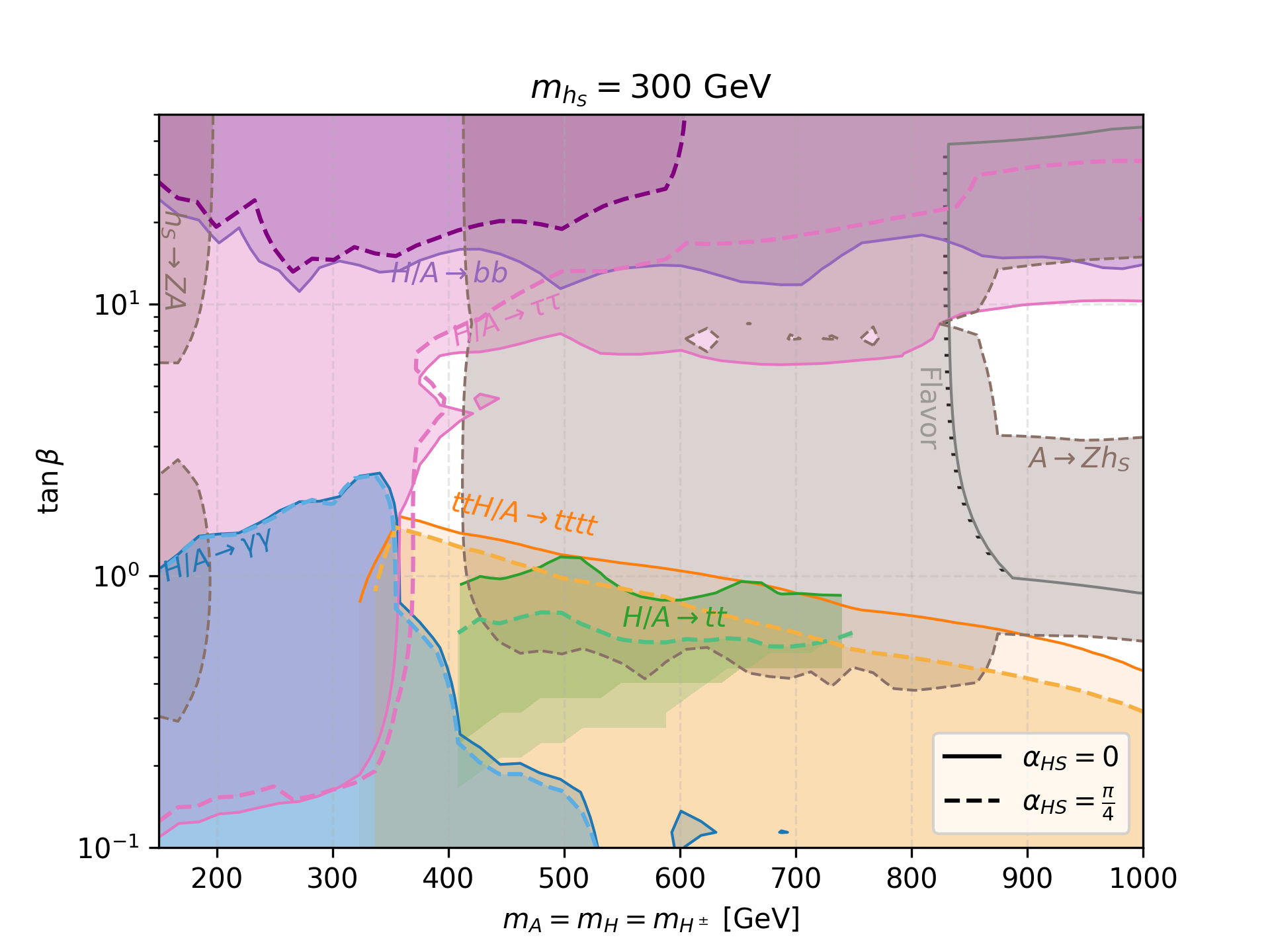}\includegraphics[width=0.5\linewidth]{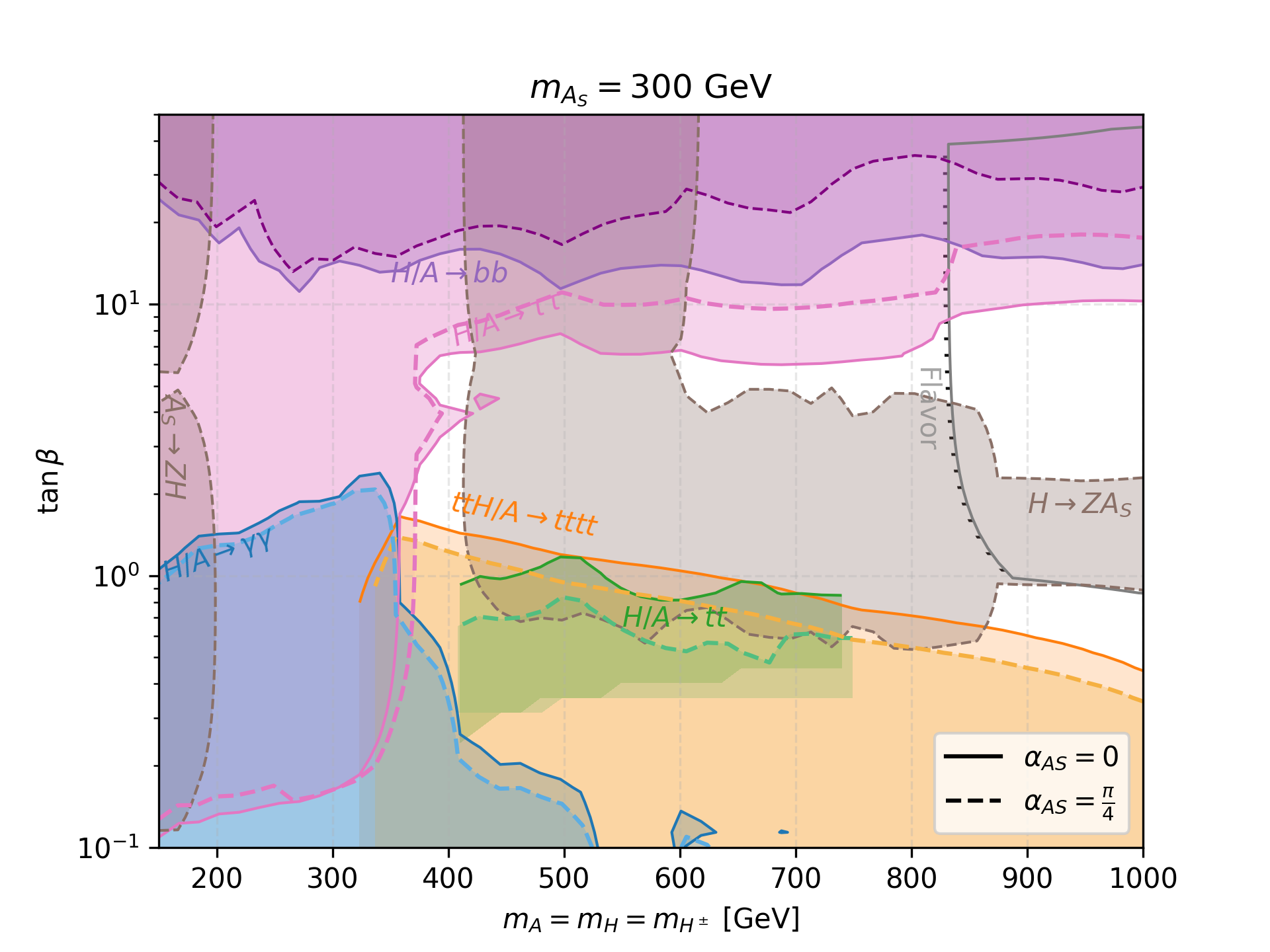}
    \caption{The 95\% C.L. exclusion region (dashed lines)   in the $\tan\beta$ vs $m_{A}=m_H=m_{H^\pm}$ plane for Case-III ($\alpha_{HS}=\frac{\pi}{4}$) and Case-IV ($\alpha_{AS}=\frac{\pi}{4}$), with the singlet Higgs mass $m_{h_S/A_S}$  set to be 300~GeV.  Case-0 is shown in solid lines for comparison. }
    \label{fig:tbma_lowms}
\end{figure}

The bottom-left panel shows the $\tan\beta$ vs. $m_{h_S}$ plane for Case-III with $\alpha_{HS}=\frac{\pi}{4}$.  
The four-top and di-top channels rule out the low $\tan\beta$ region when $m_{h_S}$ is heavier than $t\bar{t}$ threshold of 350~GeV. The large $\tan\beta$ region is mainly ruled out by $h_S\rightarrow \tau\tau$, followed by  $h_S\rightarrow bb$.  For $m_{h_S}< 350$~GeV, 
the reach of $h_S\rightarrow \tau\tau$ also extends to low $\tan\beta$ region. In this mass region, the low $\tan\beta$ region is mainly excluded by $h_S\rightarrow \gamma\gamma$.
Note that the exotic channel $A\rightarrow Z h_S $ could play an important role.  For $m_A=800$ GeV,  this channel rules out the region of low $\tan\beta$ at $m_{h_S}<550$~GeV with $A$ dominantly produced via gluon fusion.  At large $\tan\beta$ region, this channel rules out the large $\tan\beta$ region with an enhanced $bbA$ associated production. The reach gets stronger for $m_{h_S}<350$~GeV due to the enhanced $h_S\rightarrow bb/\gamma\gamma$ decay branching fraction.  
At the mass region of $m_{h_S}$ close to 125~GeV, the search of $H\rightarrow h_S h$ becomes sensitive. However, these limits of the exotic channels depend sensitively on the mass of $A$ or $H$, which is chosen to be 800 GeV for the current plot.  The reach can be enhanced (relaxed) for lighter (heavier) $A/H$.

The bottom-right panel  shows the $\tan\beta$ vs. $m_{A_S}$ plane for Case-IV with $\alpha_{AS}=\frac{\pi}{4}$. 
The $A_S\rightarrow \tau\tau$ channel excludes the region of $\tan\beta\gtrsim 0.3$ for $m_{A_S}\lesssim350$~GeV. Once above the $t\bar{t}$ threshold, $A_S\rightarrow\tau\tau$ is suppressed and only $\tan\beta \gtrsim 10$ is excluded.  
At low $\tan\beta$ region,  
$A_S\rightarrow \gamma\gamma$ is the most sensitive channel for  $m_{A_S}\lesssim350$~GeV and 
$A_S\rightarrow tt$ and $ttA_S\rightarrow tttt$ are sensitive for  $m_{A_S}\gtrsim350$~GeV. 
With $m_H$ set to be 800 GeV, $H\rightarrow Z A_S$ opens for  $m_{A_S}\lesssim 700$~GeV,  and mostly rules out the low $\tan\beta$ region.
When $m_{A_S}$ is close to 125~GeV, $A\rightarrow A_S h$ also becomes sensitive for $\tan\beta\gtrsim0.5$.

Fig.~\ref{fig:tbma_lowms} shows the 95\% C.L. exclusion regions (dashed lines)   in the $\tan\beta$ vs $m_{A}=m_H=m_{H^\pm}$ plane for Case-III ($\alpha_{HS}=\frac{\pi}{4}$) and Case-IV ($\alpha_{AS}=\frac{\pi}{4}$), with the singlet Higgs mass $m_{h_S/A_S}$  set to be 300~GeV. The zero-mixing angle case is shown in the solid lines for comparison.  The exotic channels of  $ A\rightarrow h_S Z$ and $H\rightarrow A_S Z$ open when $m_{H/A}>400$~GeV. 
For the left plot of Fig.~\ref{fig:tbma_lowms} (Case-III) with maximal mixing $\alpha_{HS}=\frac{\pi}{4}$,  $A\rightarrow Z h_S$ excludes the region of $\tan\beta>0.5$, with a small region of $4<\tan\beta<10$  still allowed at $m_A>850$~GeV due to the suppression of $A$ production cross section at the intermediate $\tan\beta$ region. For light $m_A$,  $h_S\rightarrow AZ$ exclude $m_A<200$~GeV. For the fermionic channels ($bb$, $\tau\tau$, $tt$) of the doublet Higgs $H/A$, the limits are relaxed for $\alpha_{HS}=\frac{\pi}{4}$, due to the suppressed production cross sections, as well as the suppressed branching ratios once  $A\rightarrow h_S Z$ is kinematically open. 
Similar features can be observed in the right plot of Fig.~\ref{fig:tbma_lowms} for Case-IV with $\alpha_{AS}=\frac{\pi}{4}$. The $H\rightarrow A_S Z$ limit, however, is slightly weaker due to the opening of $H\rightarrow A_S A_S$ at $m_H>600$~GeV, which suppresses the branching ratio of $H\rightarrow A_S Z$.

\subsection{$\alpha_i$ - $m_{h_i}$}
\label{sec:avm}
\begin{figure}[h]
    \centering
    \includegraphics[width=.5\linewidth]{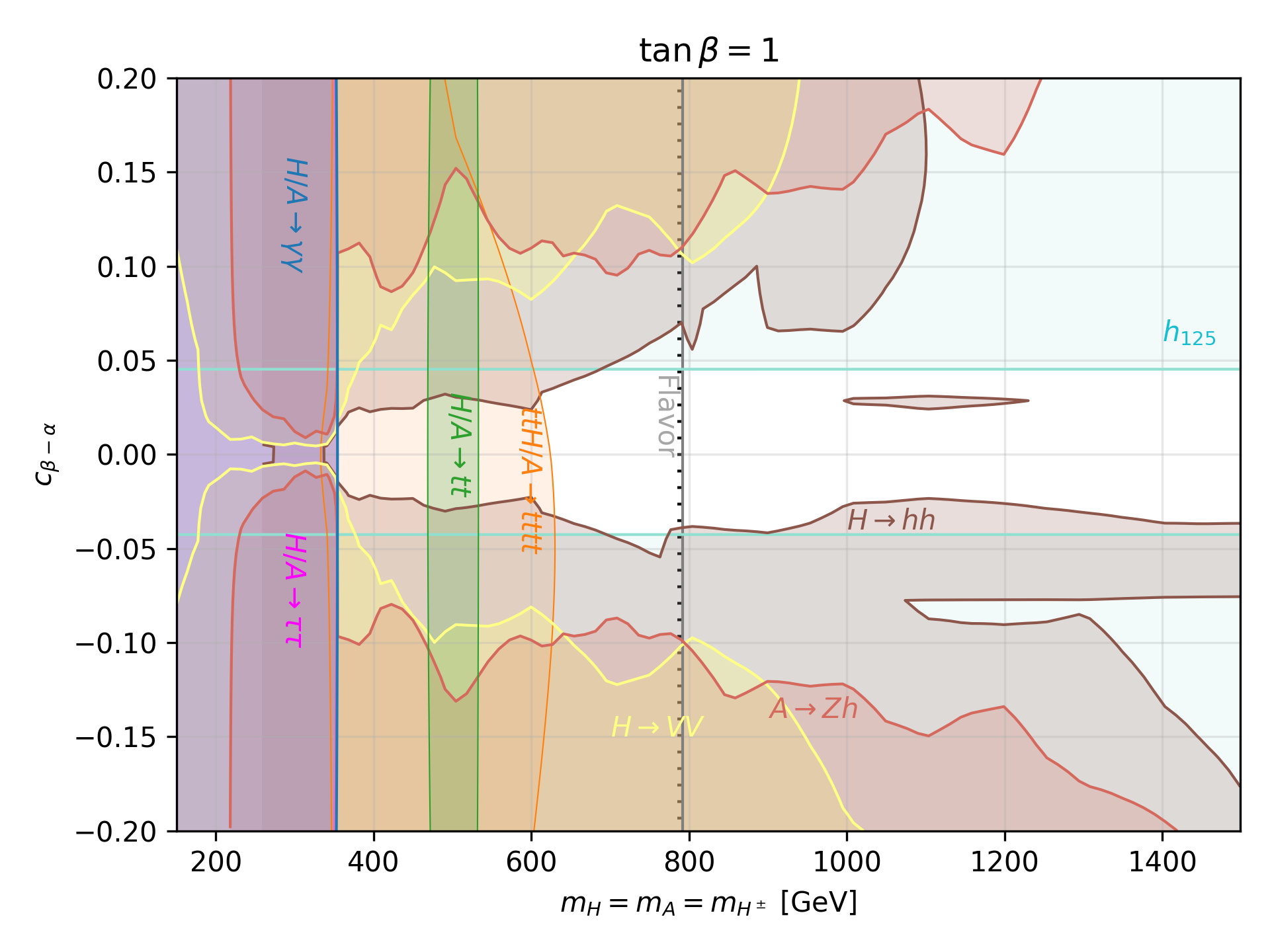}\includegraphics[width=.5\linewidth]{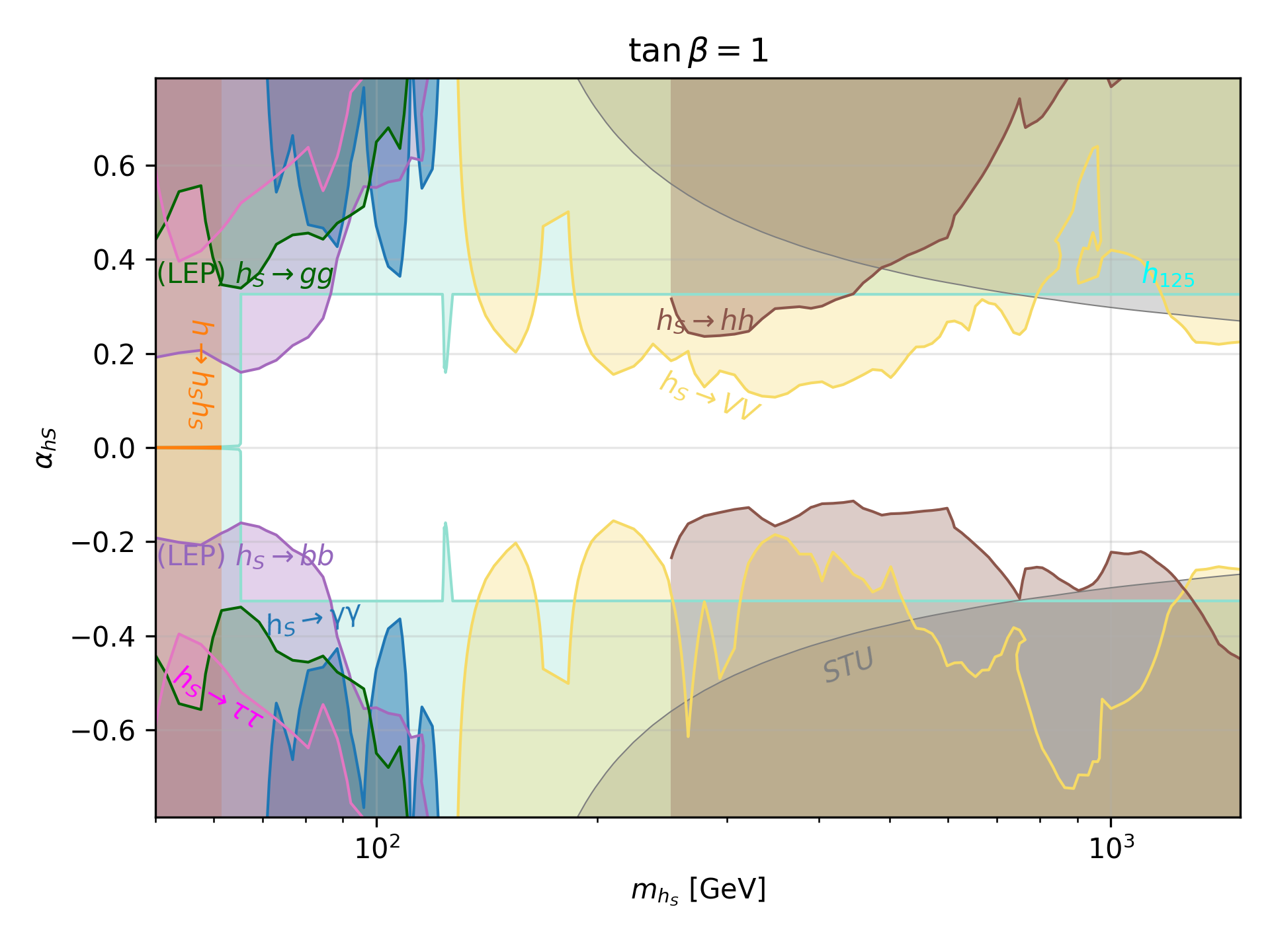}
    \includegraphics[width=.5\linewidth]{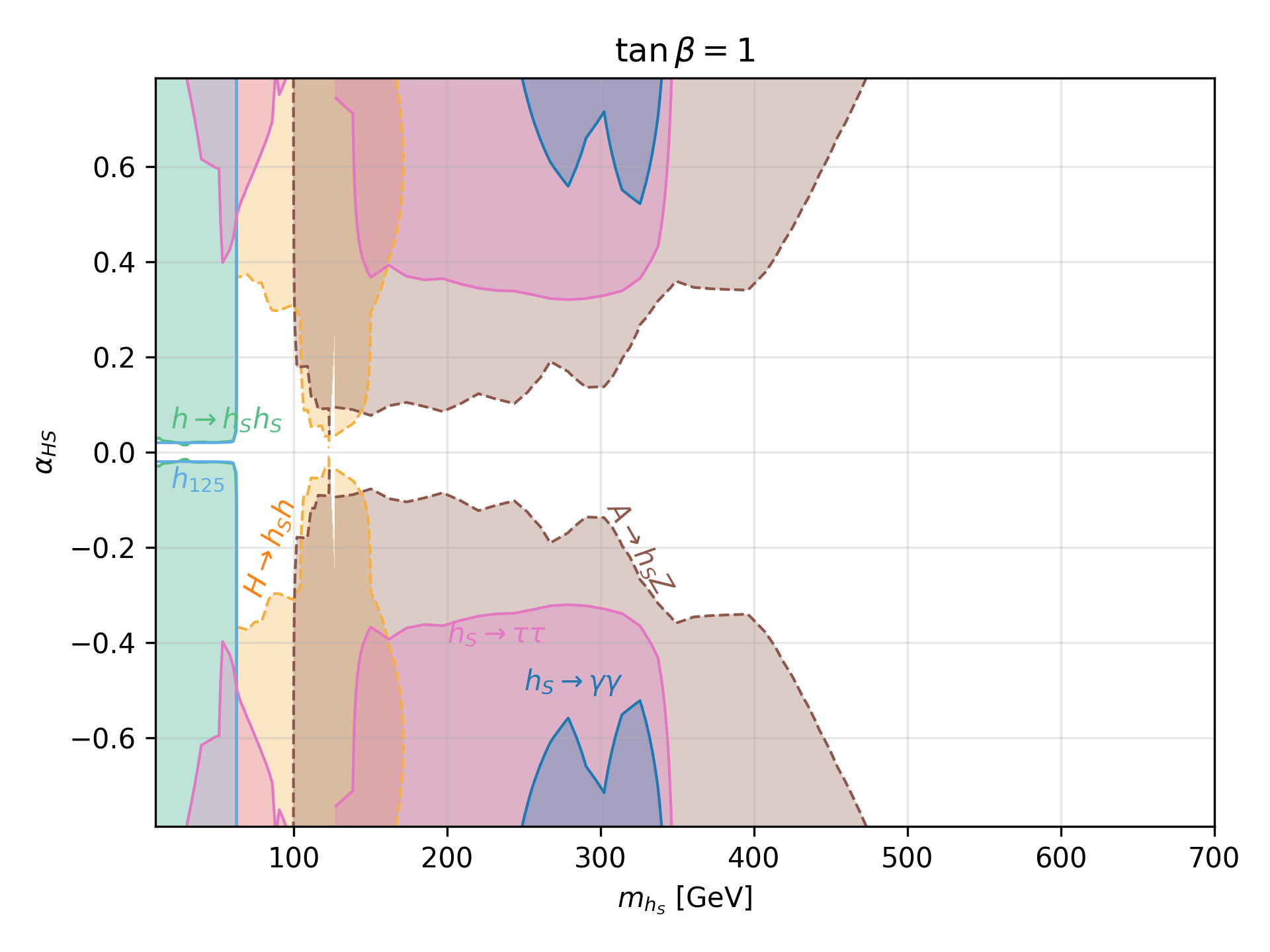}\includegraphics[width=.5\linewidth]{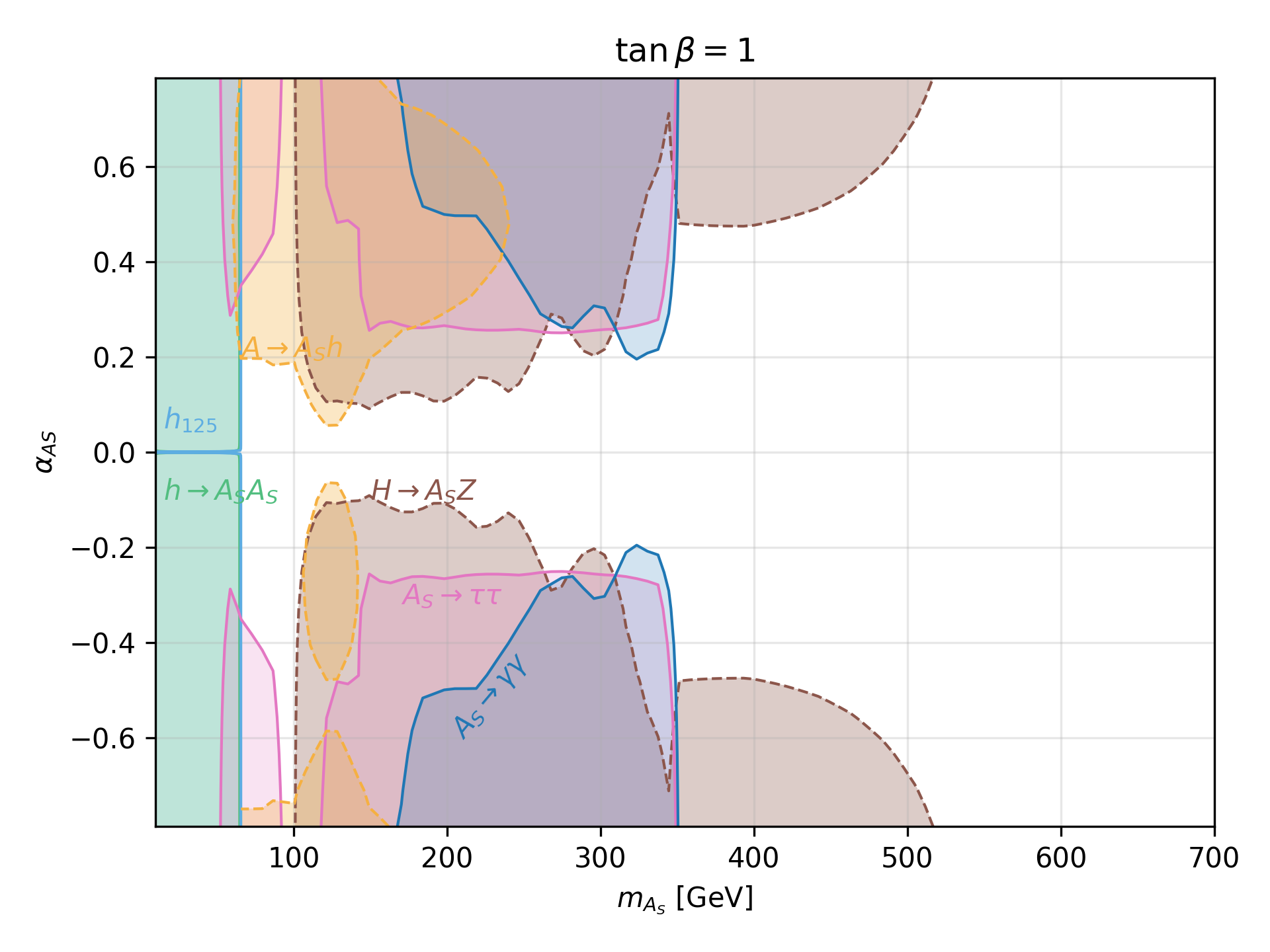}
    \caption{The 95\% C.L. exclusion regions in the parameter space of the BSM Higgs boson masses versus the corresponding mixing angles, with $\tan\beta=1$: $c_{\beta-\alpha}$ vs $m_{H/A/H^\pm}$ in  Case-I (upper left panel) with $m_{h_S}=m_{A_S}=1.5$~TeV, $\alpha_{hS}$ vs $m_{h_S}$ in  case-II (upper right panel),  $\alpha_{HS}$ vs $m_{h_S}$ in case-III (lower left panel), and   $\alpha_{AS}$ vs $m_{A_S}$ in  Case-IV (lower right panel). $m_H=m_A=m_{H^\pm}$ is set to be 800 GeV for  Case-II $-$ IV.
    }
    \label{fig:m_a}
\end{figure}

Fig.~\ref{fig:m_a} shows the 95\% C.L. exclusion regions in the parameter space of the BSM Higgs boson masses versus the corresponding mixing angles, with $\tan\beta=1$.
The upper left panel shows $c_{\beta-\alpha}$ vs degenerated doublet Higgs boson masses $m_{H/A/H^\pm}$ in  Case-I. 
The range of $c_{\beta-\alpha}$ is constrained by the SM-like $h_{125}$ coupling measurements:  $|c_{\beta-\alpha}|\lesssim0.05$. 
In addition, the di-Higgs channel $H\rightarrow hh$ plays an important role in constraining the value of $c_{\beta-\alpha}$. The asymmetric constraints on the sign of $c_{\beta-\alpha}$ is due to the asymmetric $Hhh$ coupling. 
The four-top and di-top channels cover $350~ {\rm GeV}<m_{H/A}<600$ GeV, while 
$H/A\rightarrow\gamma\gamma$ and $H/A\rightarrow\tau\tau$ (two lines overlap) mainly exclude the $m_{H/A}<350$ GeV region, even for $c_{\beta-\alpha}=0$. 
In addition, the $H\rightarrow VV$ and $A\rightarrow Z h$ channels become sensitive when away from the alignment limit, especially below the $t\bar{t}$ threshold.

The upper right panel of Fig.~\ref{fig:m_a} shows the parameter space of the $\alpha_{hS}$ vs CP-even singlet Higgs bosons mass $m_{h_S}$ in    Case-II.
For $m_{h_S}\gtrsim 125$~GeV, the dominant limit of $\alpha_{hS}$ comes from the $h_S\rightarrow VV$ and $h_S\rightarrow hh$. Due to the asymmetric nature of $h_S\rightarrow hh$ on the sign of $\alpha_{hS}$, $h_S\rightarrow hh$ is more sensitive for  $\alpha_{hS}<0$, while $h_S\rightarrow VV$   is more sensitive for  $\alpha_{hS}>0$. In addition, the $STU$ constraint becomes sensitive for heavy $m_{h_S}$~\cite{Li:2025zga}.
For 62.5 GeV $\lesssim m_{h_S}\lesssim125$~GeV, the $\alpha_{hS}$ limit is dominated by the LEP constraints of $e^+e^-\rightarrow Zh_S\rightarrow Zbb$. For $m_{h_S}\lesssim 62.5$~GeV, $h \rightarrow h_S h_S$ rules out the entire region of $\alpha_{hS}$ unless $\alpha_{hS}$ is very close to zero.

The lower left panel of Fig.~\ref{fig:m_a} shows the parameter space of the $\alpha_{HS}$ vs CP-even singlet Higgs bosons mass $m_{h_S}$ in Case-III. $A\rightarrow h_S Z$ and $H\rightarrow h_S h$ are the most sensitive channels for $h_S$ searches with $m_{hS}>62.5$ GeV.  These two limits, however, depend sensitively on the value of $m_{H/A}$, which is chosen to be 800 GeV for this plot.  In the lighter mass region, the $h \rightarrow h_Sh_S$ strongly constrains the $\alpha_{HS}$. Both $h$ precision measurement and direct search of $h\rightarrow h_S h_S$ yield constraints of $|\alpha_{HS}|\lesssim0.02$ at the low mass region.  
Since $h_S$ only mixes with $H$ in this case,   $h_SVV$ coupling is absent and the LEP limits for CP-even Higgs do not apply.
Note that for  $\tan\beta=1$, the branching ratio of  $h_S\rightarrow tt$ is not large enough to be constraining.

The lower right panel of Fig.~\ref{fig:m_a} shows the parameter space of the $\alpha_{AS}$ vs CP-odd singlet Higgs bosons mass $m_{A_S}$ in  Case-IV. 
The $H\rightarrow A_S Z$ channel provides the strongest limit for $m_{A_S}\gtrsim 150$~GeV, followed by $A_s\rightarrow \tau\tau, \gamma\gamma$. For $100~\mathrm{GeV}\lesssim m_{A_S}\lesssim 150$~GeV, the $A\rightarrow A_S h$ channel becomes effective.  Both limits depend sensitively on the value of $m_{H/A}$ chosen.    Similar to the other cases, the lighter region $m_{A_S}\lesssim 62.5$~GeV would be mostly excluded by $h_{125}$ measurement and $h\rightarrow A_S A_S$ decay, unless $\alpha_{AS}$ is very close to zero.

\begin{figure}[h]
    \centering
    \includegraphics[width=0.5\linewidth]{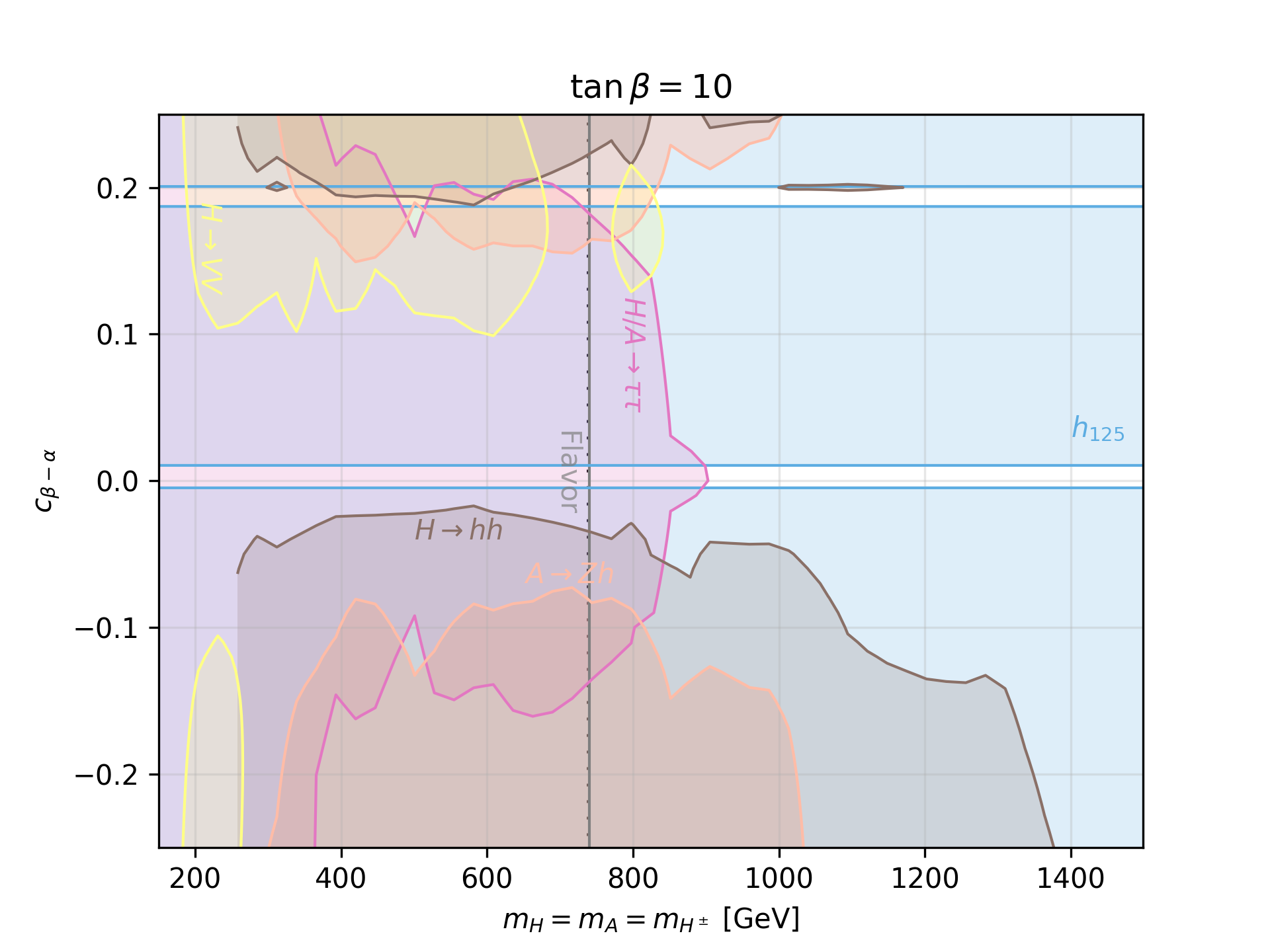}\includegraphics[width=0.5\linewidth]{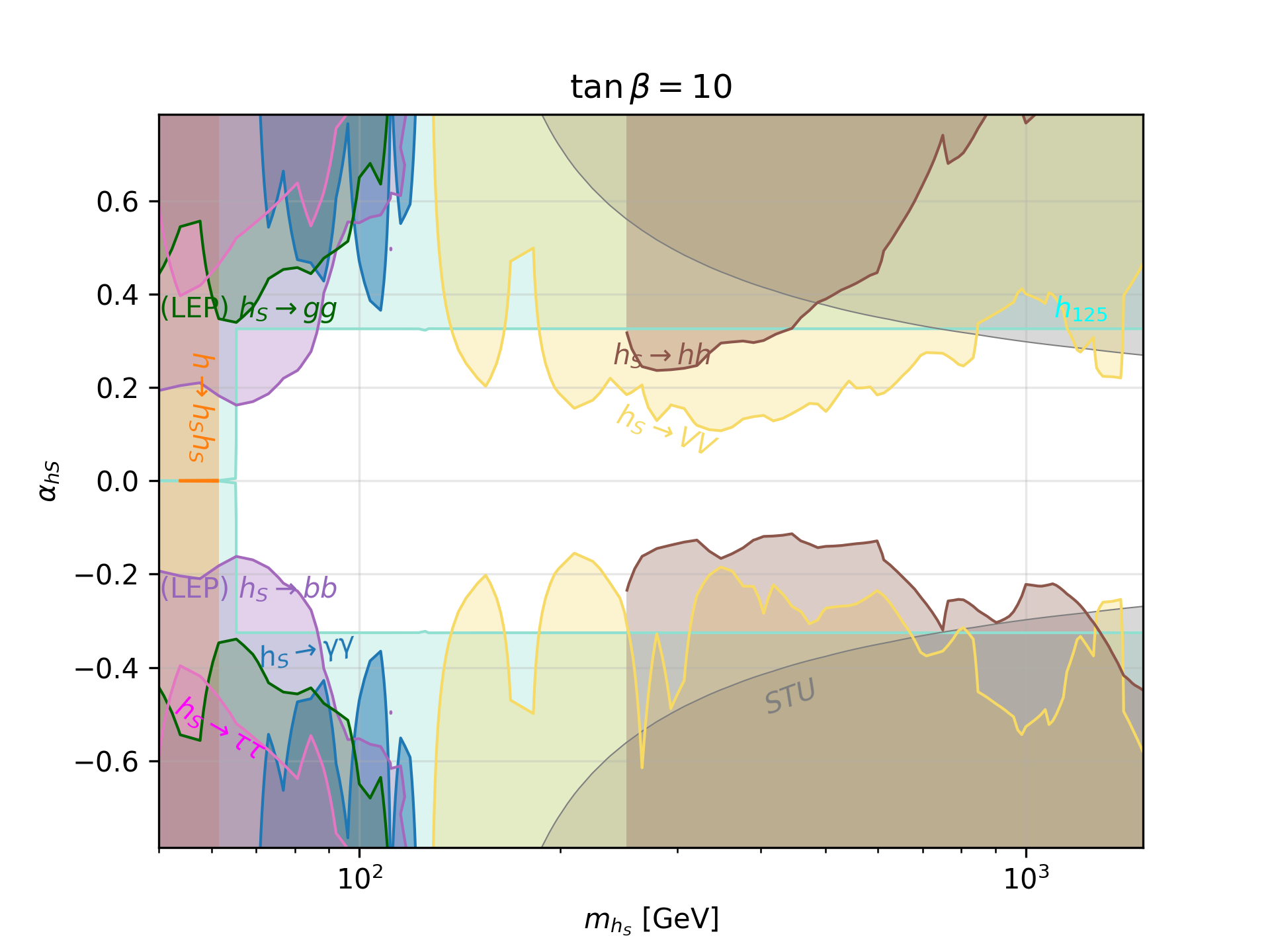}
    \includegraphics[width=0.5\linewidth]{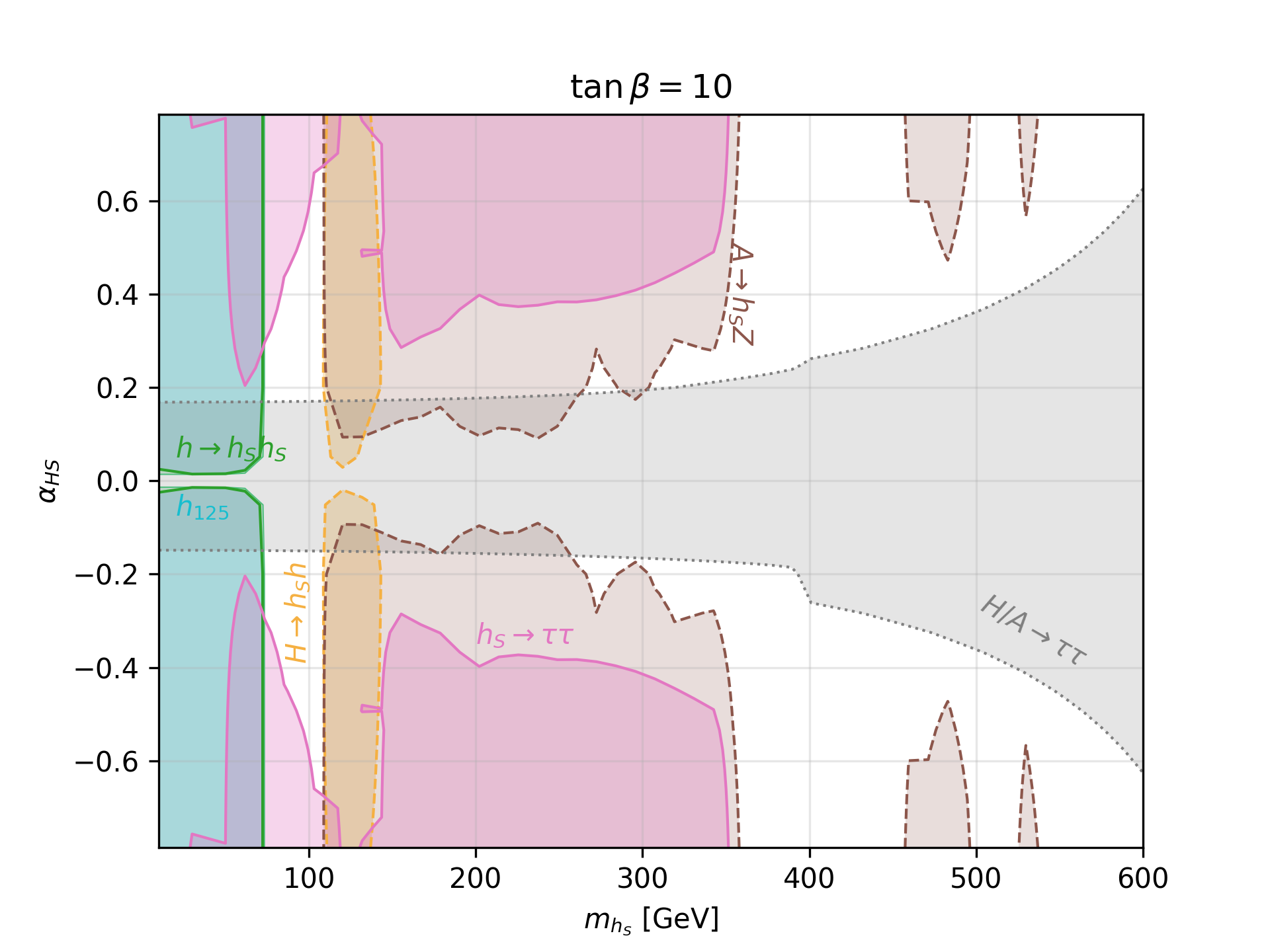}\includegraphics[width=0.5\linewidth]{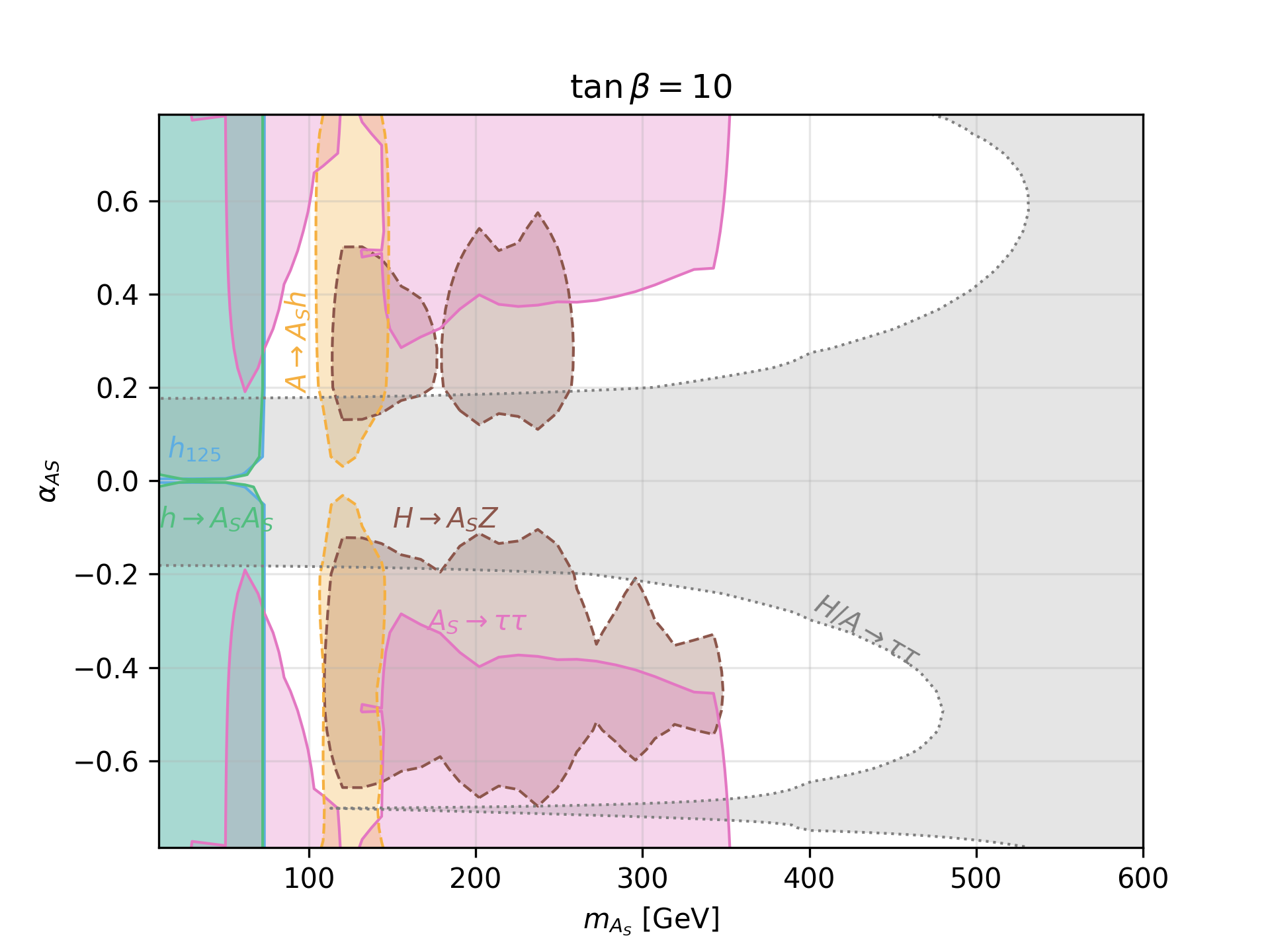}
    \caption{The 95\% C.L. exclusion regions in the parameter space of BSM Higgs boson masses
versus the corresponding mixing angles similar to Fig.~\ref{fig:m_a}, but with   $\tan\beta=10$. }
    \label{fig:m_a_tb10}
\end{figure}

To illustrate the dependence on $\tan\beta$, in Fig.~\ref{fig:m_a_tb10}, we show the 95\% C.L. exclusion regions in Higgs boson mass vs. mixing angle parameter space for $\tan\beta=10$. For   Case-I (the upper plot), the SM-like Higgs coupling precision measurements tightly constrain the range of  $|c_{\beta-\alpha}|$ to be less than 0.01.  Note that $0.18\lesssim c_{\beta-\alpha} \lesssim 0.2$ is still allowed as the ``wrong sign" region~\cite{Su:2019ibd}.  $H/A\rightarrow \tau\tau$ provides the dominant constraints for $m_{A,H} <$ 900 GeV. The di-Higgs channel $H\rightarrow hh$ at negative $c_{\beta-\alpha}$ region is enhanced while $H\rightarrow VV$ is suppressed. At the positive $c_{\beta-\alpha}$ region, the $H\rightarrow hh$ is much more suppressed, and $H\rightarrow VV$ becomes more constraining.

For  Case-II, the behaviour of all the channels shows weak dependence on $\tan\beta$, as shown in the top right panel of Fig.~\ref{fig:m_a_tb10}.  In this case, only the $h_S\rightarrow hh$ limit depends slightly on $\tan\beta$, due to the trilinear Higgs coupling $g_{h_S hh}$. Consequently, the $h_S \rightarrow VV$ branching ratio is also affected and demonstrates a small dependence on $\tan\beta$.  

For  Case-III (lower left panel), $h_S\rightarrow\gamma\gamma$ is no longer sensitive, since the $h_Stt$ coupling is highly suppressed. 
$A \rightarrow h_S Z$ provides the dominant constraints for $m_{h_S} \gtrsim 100$ GeV.  There is a gap between 350 $-$ 450 GeV, due to the statistical fluctuation in the limit of the experimental search channel $bbA\rightarrow Z h_S, h_S\rightarrow bb$~\cite{ATLAS:2011.05639}. 
For small $\alpha_{HS}$, the parameter space is  excluded by $H/A\rightarrow \tau\tau$ for $m_H=m_A=800$ GeV. Such constraint is relaxed for larger $|\alpha_{HS}|$ given the suppressed $H/A\rightarrow\tau\tau$. For larger $m_{h_S}$ with fixed $\alpha_{AS}$, the branching fraction of $H/A \rightarrow \tau\tau$ is larger given the suppression of $A\rightarrow Z h_S$ and $H\rightarrow h h_S$, which leads to stronger constraints for $H/A\rightarrow\tau\tau$ channel.

For  Case-IV (lower right panel), the behavior of $H/A \rightarrow\tau\tau$ channels (gray region) is similar to Case-III, except that the limits get weaker when $|\alpha_{AS}|\sim \pi/8$ or $\pi/4$. This is because the $A\rightarrow A_S h$ becomes dominant for $|\alpha_{AS}|\sim \frac{\pi}{8}$ or $\pi/4$, given that the coupling $g_{hA A_S}$ receives two contributions, one $\propto \sin(4\alpha_{AS})$, and the other $\propto \sin(2\alpha_{AS})$ (see Table~\ref{tab:haamp}). When $|\alpha_{AS}|$ deviates away from these two values, the $A\rightarrow A_S h$ decay width is suppressed, and the $A\rightarrow\tau\tau$ increases, which leads  to tighter constraints. 

\subsection{$\tan\beta$ - $\alpha_i$}
\label{sec:tbva}
\begin{figure}[h]
    \centering
    \includegraphics[width=.5\linewidth]{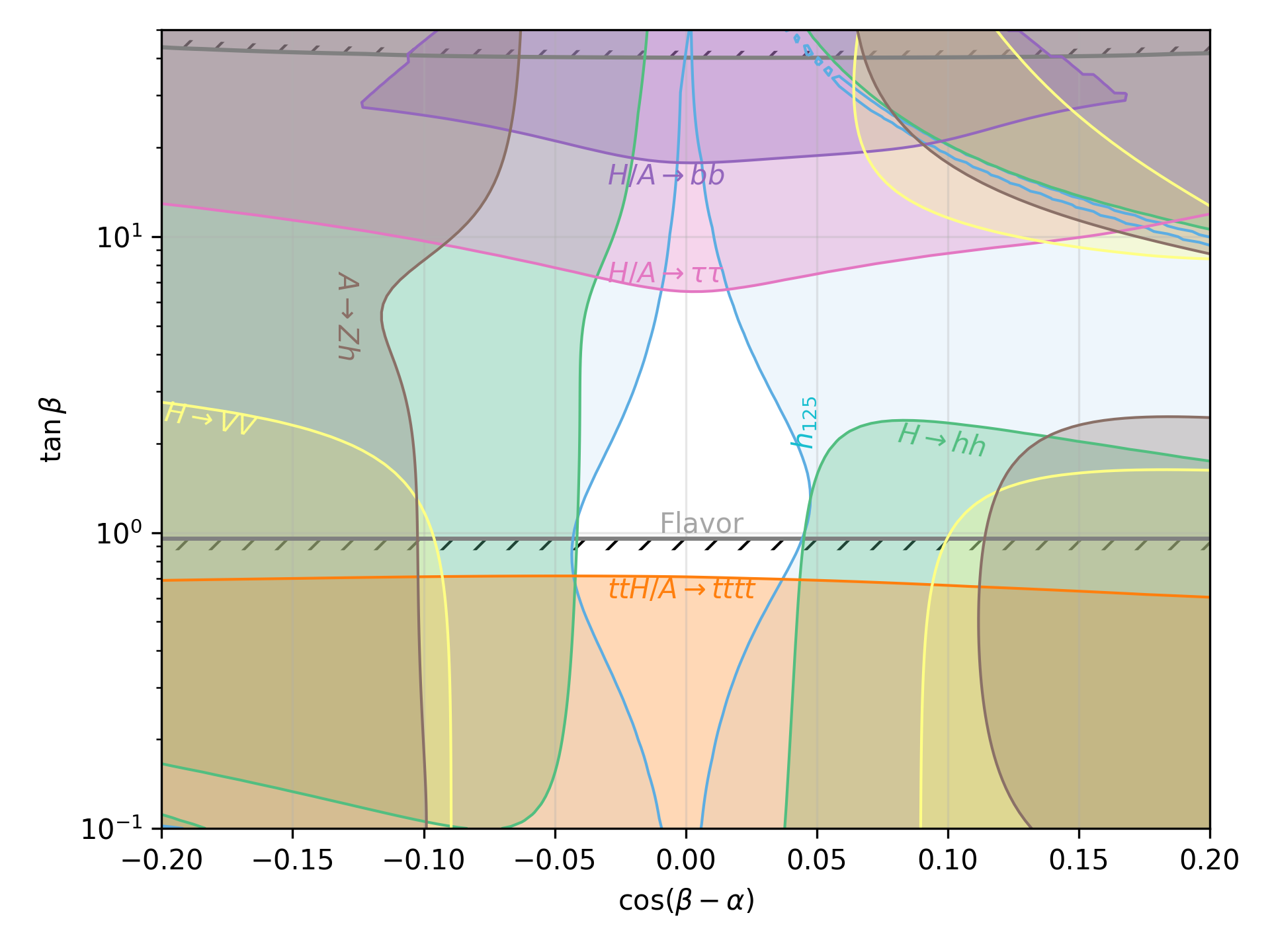}\includegraphics[width=.5\linewidth]{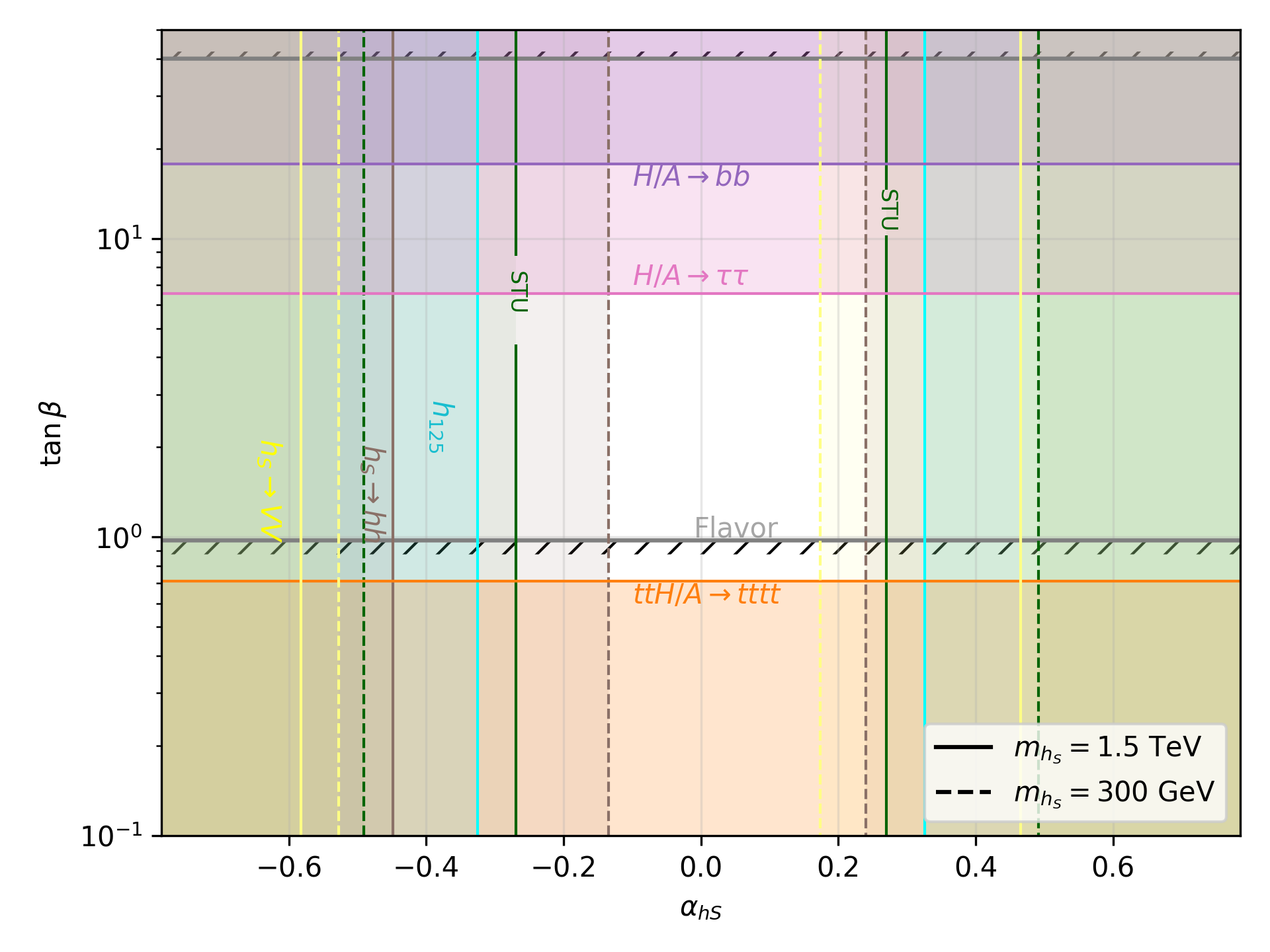}
    \includegraphics[width=.5\linewidth]{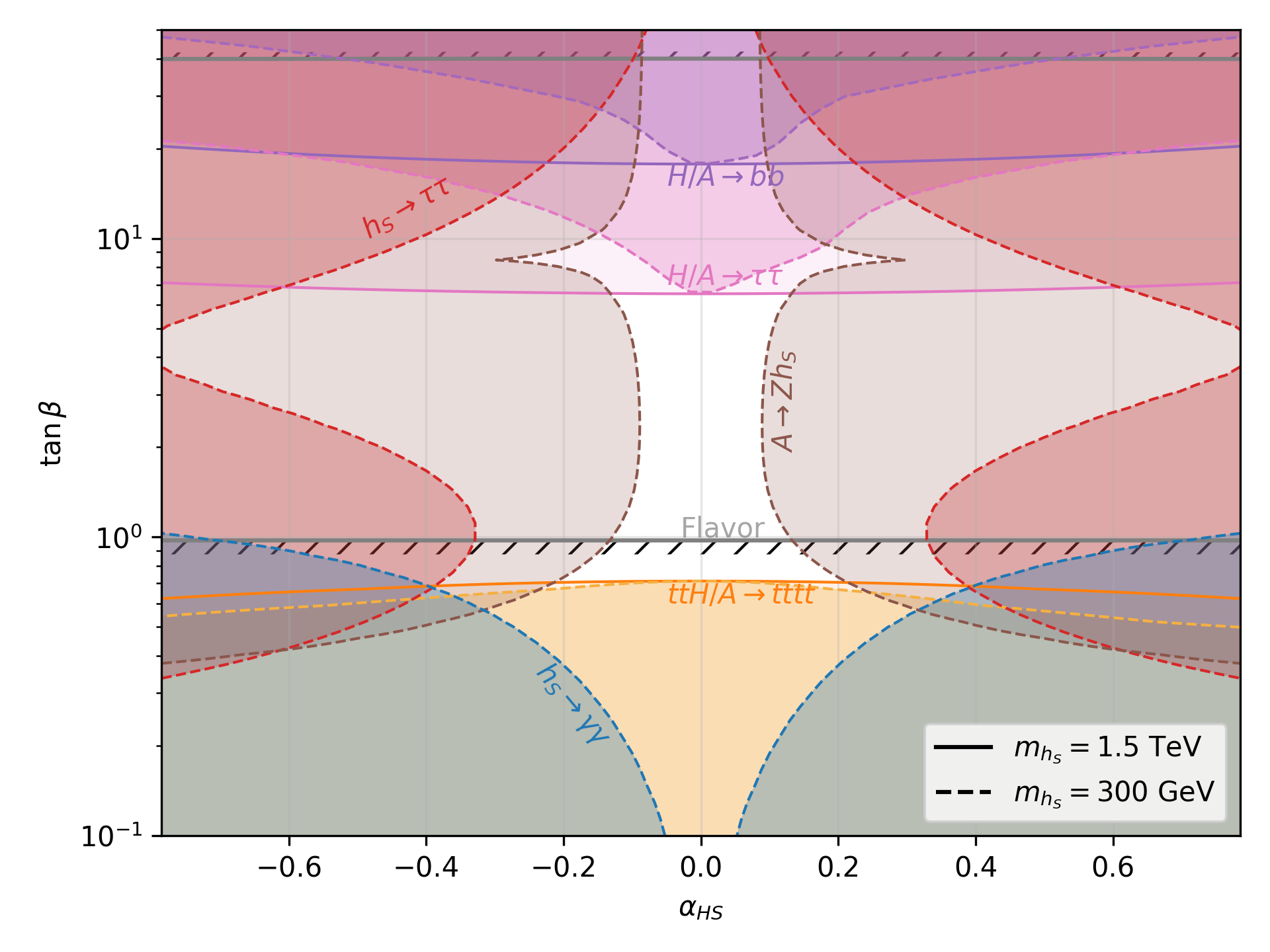}\includegraphics[width=.5\linewidth]{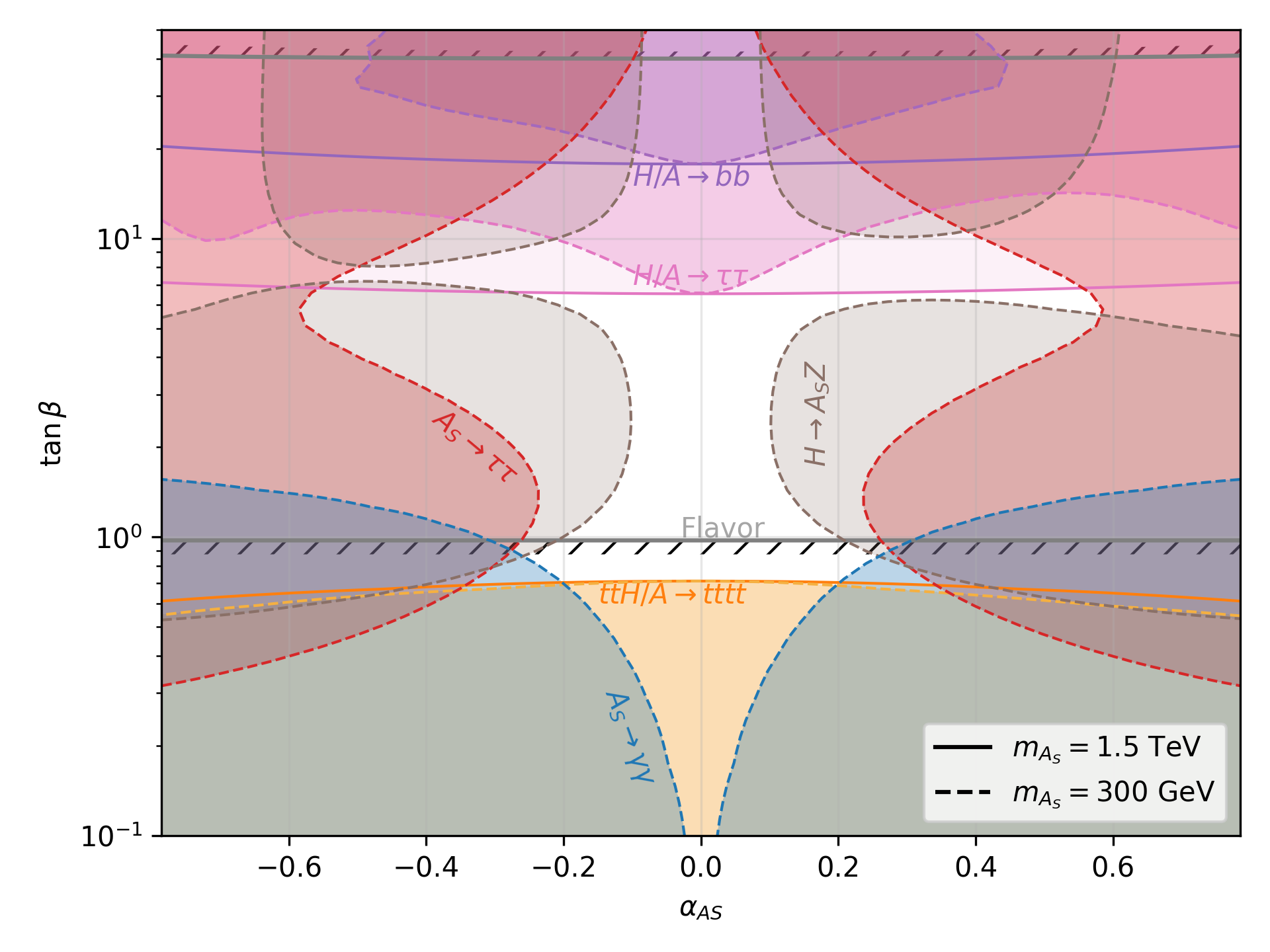}
    \caption{The 95\% C.L. exclusion regions in the parameter space of $\tan\beta$ vs mixing angles with $m_H = m_A = m_{H^\pm}=m_{\phi}=800$~GeV and $v_S = v$. The upper left panel is for the Case-I ($\tan\beta$ vs $c_{\beta-\alpha}$) with $m_{h_S}=m_{A_S}=1.5$~TeV. The upper right panel is for the Case-II ($\tan\beta$ vs $\alpha_{hS}$) and the lower left panel is for the Case-III ($\tan\beta$ vs $\alpha_{HS}$)  for $m_{h_S}=1.5$~TeV (solid lines) and $m_{h_S}=300$~GeV (dashed lines).
    The lower right panel is for the Case-IV ($\tan\beta$ vs $\alpha_{AS}$),  for $m_{A_S}=1.5$~TeV (solid lines) and $m_{A_S}=300$~GeV (dashed lines). }
    \label{fig:tb_a}
\end{figure}

In Fig.~\ref{fig:tb_a}, we present the 95\% C.L. exclusion regions in the $\tan\beta$ vs  mixing angle parameter space, with four panels representing Case-I $-$ IV. 
In the upper left panel (Case-I),  the 125~GeV Higgs coupling precision measurements mainly constrain the $\cos(\beta-{\alpha})$ to be around that alignment limit, which is consistent with the 2HDM results~\cite{Gu:2017ckc}. The region with large $\tan\beta$ is excluded by direct searches of $H/A\rightarrow \tau^+\tau^-, bb$, while the region with small  $\tan\beta$ is constrained by the four-top searches $t\Bar{t}H/A\rightarrow \bar{t}\bar{t}tt$.
Region with large $|\cos(\beta-\alpha)|$ is excluded by  $H\rightarrow VV$ and $A\rightarrow Z h$ since the $HVV$ and $AhZ$ couplings are  proportional to   $\cos(\beta-\alpha)$.  The relaxing of $H\rightarrow VV$ at intermediate $\tan\beta$ is due to the suppression of the production cross sections.
Note that the wrong sign region of the Higgs precision measurements on the top right can be excluded by $H\rightarrow VV$ and $A\rightarrow Z h$ for $m_{H/A}=800$~GeV, while the heavier ${H/A}$ leads to weaker constraints, as shown in the upper left panel of Fig.~\ref{fig:m_a_tb10}.
The di-Higgs channel of $H \rightarrow hh$ opens for $m_H = 800$~GeV. The limits are stronger in the negative $\cos({\beta-\alpha})$ region due to the larger trilinear $\lambda_{Hhh}$ coupling for $\cos({\beta-\alpha})<0$, which is consistent with the upper left plot of Fig.~\ref{fig:m_a}.

In the upper right panel (Case-II), the constraints from the doublet Higgs $H/A\rightarrow \tau\tau$, $tt$ and $bb$ are independent of $\alpha_{hS}$, while the constraints from the singlet $h_S\rightarrow VV$ and $hh$ are independent of the $\tan\beta$. $h_S$ based constraints also highly depend on the value of $m_{h_S}$, which are shown in solid (dashed) vertical lines for $m_{h_S}=1.5$ TeV (300 GeV). Limits from the $h_S\rightarrow VV$ and $hh$ channels get considerably stronger for smaller $m_{h_S}$, which can also be found in upper right plot of Fig.~\ref{fig:m_a} (Case-II). 
Note that other $h_S$ decay channels of $\gamma\gamma$, $bb$, and $\tau\tau$  are too weak to constrain the parameter space for $m_{h_S}=300$ GeV.

In the lower left panel (Case-III) with a heavy singlet ($m_{h_S}=1.5$~TeV, solid curves), $\tan\beta$ receives the usual constraints from $H/A$ with very weak dependence on $|\alpha_{HS}|$.  For a light singlet $m_{h_S}=300$~GeV, the constraints from $H/A \rightarrow tt$, $bb$ and $\tau\tau$ are weakened at large $|\alpha_{HS}|$ due to the opening of  $A \rightarrow Z h_S$. This channel, by itself, also imposes strong constraints on $\alpha_{HS}$ for $\tan\beta > 0.4$. Furthermore, $h_S\rightarrow \tau\tau$ and $h_S\rightarrow\gamma\gamma$ kick in at large and small $\tan\beta$ as well.  The relaxing of $h_S\rightarrow \tau\tau$ at $\tan\beta \sim$ 4 is due to the suppression of the production cross sections. 

In the lower right panel (Case-IV) the mixture of CP-odd singlet $A_S$ with 2HDM $A$, the characteristic of the plot is very similar to Case III with the substitution of $(h_S, \alpha_{hS})$ with  $(A_S, \alpha_{AS})$. The reach of $H\rightarrow A_S Z$  is reduced in $\tan\beta\sim 8$ due to suppression of production cross sections. The constraints from the $H\rightarrow Z A_S$ channel is relaxed at  $|\alpha_{AS}|\sim\frac{\pi}{4}$ and large $\tan\beta$. This is because in this region, $g_{HA_S A_S}\sim \sin^4\alpha_{AS}/\tan2\beta$ (see Table~\ref{tab:haamp}) is enhanced, and  $H\rightarrow A_S A_S$ decay becomes dominant.

\begin{figure}
    \centering
    \includegraphics[width=0.5\linewidth]{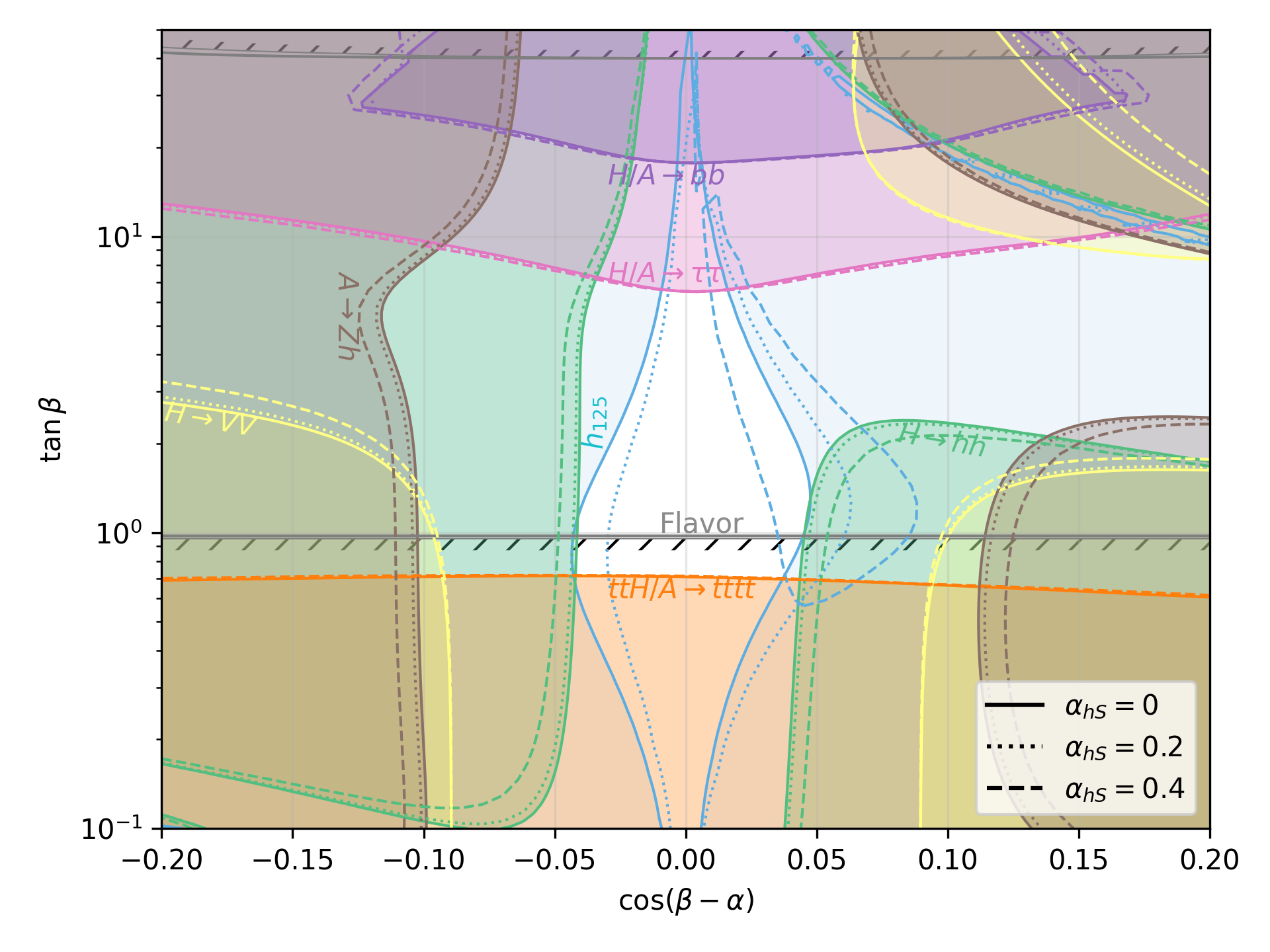}\includegraphics[width=0.5\linewidth]{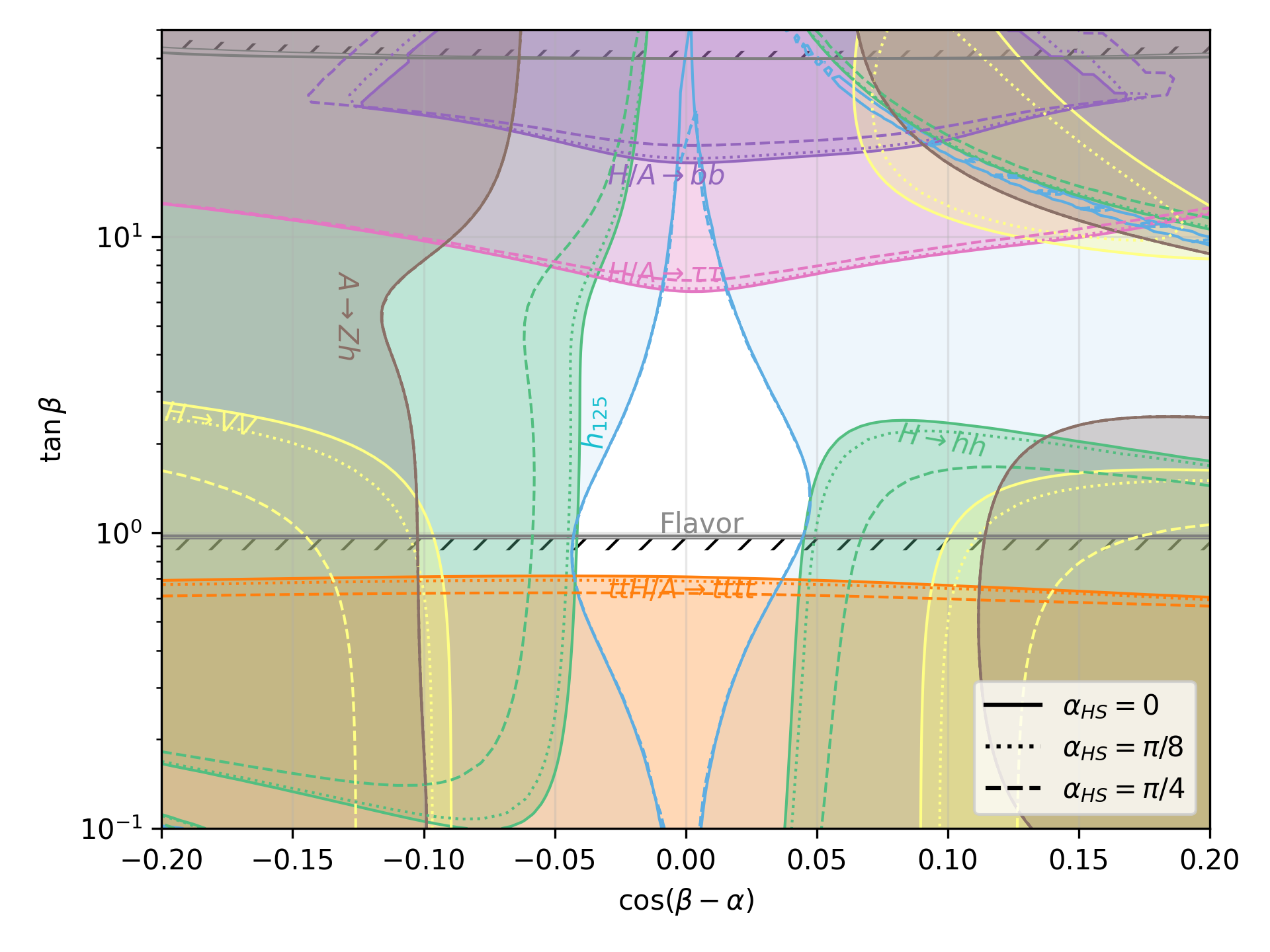}
    \caption{The 95\% C.L. exclusion regions in  $\tan\beta$ vs $c_{\beta-\alpha}$ plane with $m_H = m_A = m_{H^\pm}=m_{\phi}=800$~GeV, $m_{h_S}=m_{A_S}=1.5$~TeV and $v_S = v$. The solid, dotted and dashed lines in the left (right) panel are for $\alpha_{hS} =$0, 0.2 and 0.4 ($\alpha_{HS} =0, \pi/8$ and $\pi/4$), respectively. }
    \label{fig:tb_cba_a}
\end{figure}
In Fig.~\ref{fig:tb_cba_a}, we show the 95\% C.L. exclusion regions in the $\tan\beta$ vs $c_{\beta-\alpha}$ plane with varying $\alpha_{hS}$ and $\alpha_{HS}$. In the left plot, increasing $\alpha_{hS}$ would shift the allowed parameter space of the 125~GeV Higgs boson precision measurement to the positive $c_{\beta-\alpha}$ region. In particular for  $\alpha_{hS}=0.4$, $c_{\beta-\alpha}=0$ is excluded, and the maximal allowed value for $c_{\beta-\alpha}$ can reach 0.08 for $\tan\beta\sim 1$. The di-Higgs channel $H\rightarrow hh$ and  $A\to Z h $  are suppressed at larger $\alpha_{hS}$ due to the reduced couplings. $H\to VV$, $bb$, $\tau\tau$ and $tt$, on the other hand, are  slightly enhanced.  

In the right plot of Fig.~\ref{fig:tb_cba_a} with varying $\alpha_{HS}$, the 95\% C.L. exclusion regions of the 125~GeV Higgs boson measurement and $A\to Z h$ channel are  independent of $\alpha_{HS}$.  All other direct search channels of  $H\to VV,~bb,~\tau\tau, tt$ and $hh$ are slightly suppressed  with increasing $\alpha_{HS}$ due to the reduced  production rate of $H$.

\subsection{$m_{h_i}$ - $m_{h_j}$}
\label{sec:mvm}
\begin{figure}[h]
    \centering
    \includegraphics[width=.5\linewidth]{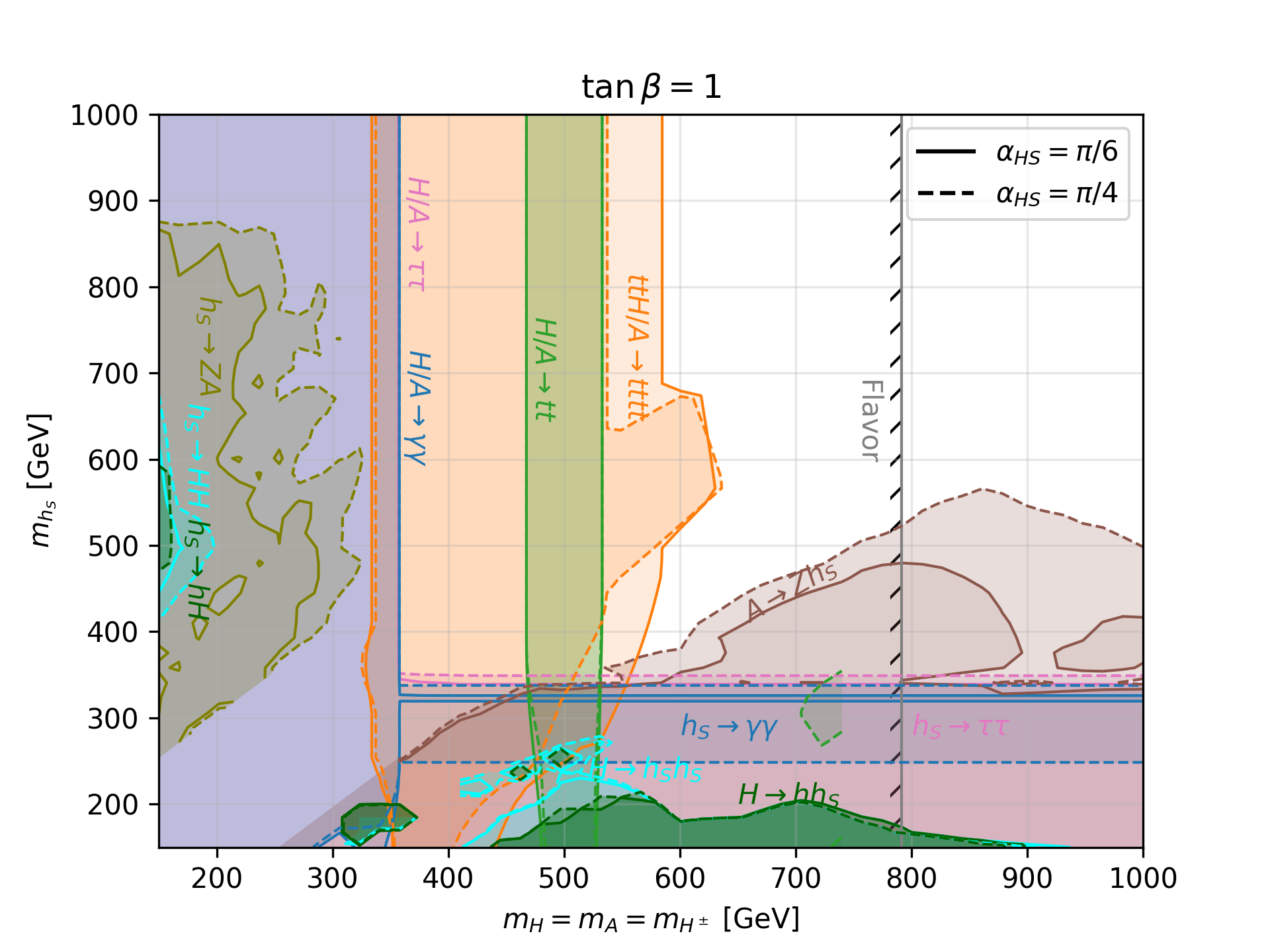}\includegraphics[width=.5\linewidth]{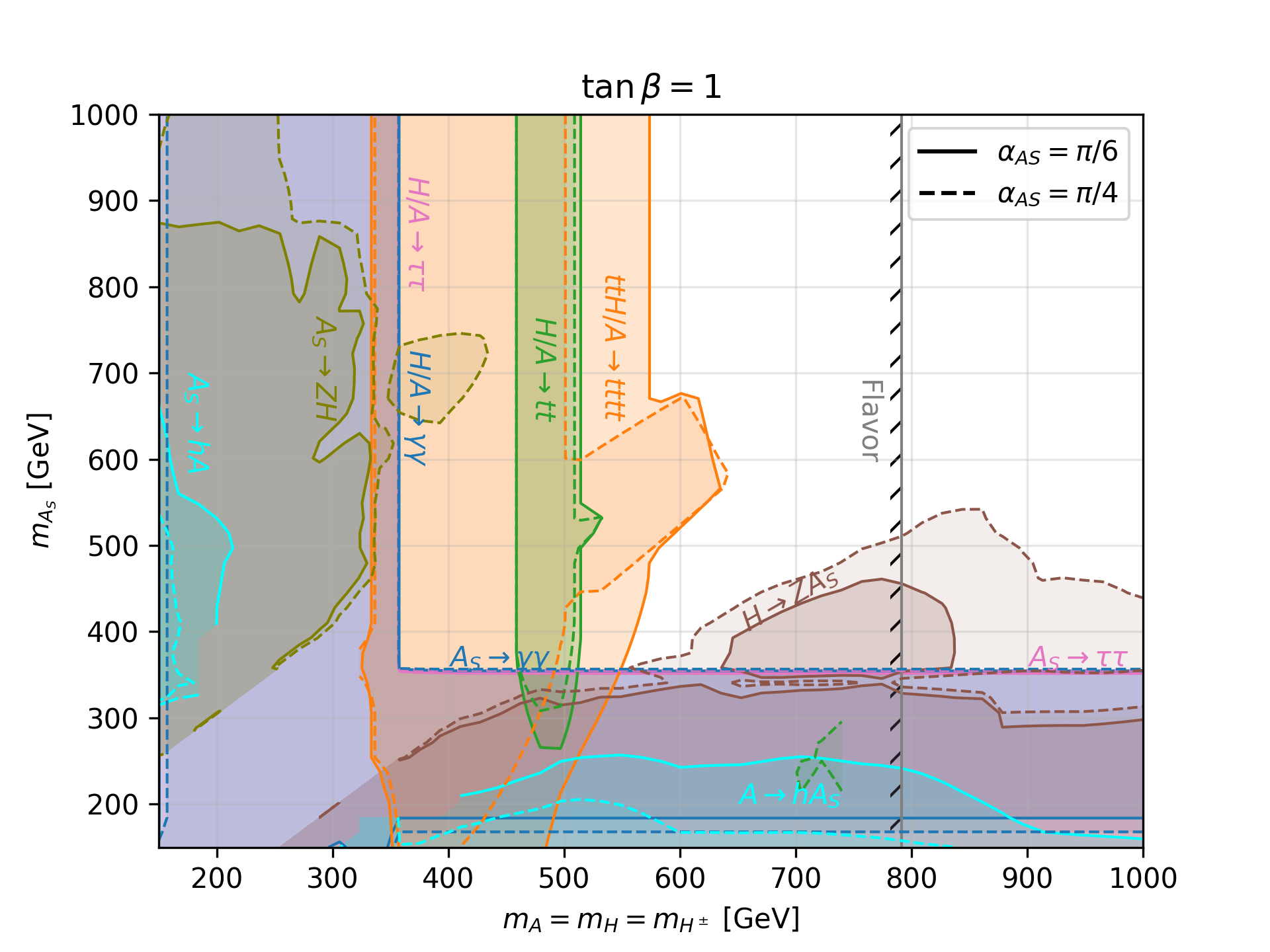}
    \caption{The 95\% C.L. exclusion regions of the doublet Higgs boson masses $m_{H/A/H^{\pm}}$ vs singlet Higgs boson mass $m_{h_S}$ in Case-III (left panel) and 
    $m_{A_S}$ in Case-IV (right panel).  The solid and dashed curves are for $\alpha_{HS, AS}=\pi/6$ and  $\pi/4$ respectively.  Both panels have $\tan\beta=1$ and $v_S=v$.}
    \label{fig:mh-ms}
\end{figure}
The left panel of Fig.~\ref{fig:mh-ms} shows the 95\% C.L. exclusion region of $m_{H/A/H^{\pm}}$ vs $m_{h_S}$ in  Case-III. For $m_{H/A/H^\pm}\lesssim 550$~GeV and $m_{H/A/H^\pm}<m_{h_S}$, the parameter space is excluded by the four-top channel, where the diagonal region of $m_{h_S}\sim m_{H/A/H^\pm}$ has an enhancement, since the contribution of $h_S$ and $H/A$ are combined. For $m_{h_S} \lesssim m_{H/A/H^\pm}$, the limits from the four top channel is weakened due to the opening of $H$ and $A$ decay with $h_S$ in the final states. The di-top $H/A\rightarrow tt$ channel rules out the region $m_{H/A/H^\pm}\sim 500$~GeV and is relatively insensitive to $m_{h_S}$. The region of  $m_{H/A/H^\pm}\lesssim$350~GeV is ruled out by the $H/A\rightarrow \tau\tau$ and $\gamma\gamma$ channels.  $h_S\rightarrow\tau\tau$ also  excludes $m_{h_S}\lesssim350$~GeV. The limit of $h_S\rightarrow \gamma\gamma$ channel depends on the value of $\alpha_{HS}$: $250\lesssim m_{h_S}\lesssim350$~GeV is excluded for $\alpha_{HS}=\pi/4$, while the exclusion region is suppressed to a small region of $m_{h_S}\sim 350$~GeV for $\alpha_{HS}=\pi/6$. For $m_{h_S} \lesssim m_{H/A/H^\pm}$ at $\tan\beta=1$, $A\rightarrow Z h_S$ opens and becomes the dominant constraint at large $m_{h_S}$. For $\alpha_{HS}=\pi/4$, $m_{h_S}$ up to about 550 GeV is ruled out for $m_{A}$ up to about 850~GeV.  The limits get slightly weaker for $\alpha_{HS}=\pi/6$. 

The right panel of Fig.~\ref{fig:mh-ms} shows the 95\% C.L. exclusion region of $m_{H/A/H^{\pm}}$ vs $m_{A_S}$ in   Case-IV. The features are similar to those of the $h_S-h$ mixing case in the left panel, with the exchange of $(h_S, \alpha_{HS})$ with  $(A_S, \alpha_{AS})$. The limit from $A_S\rightarrow \gamma\gamma$ is stronger and excludes the whole region of $m_{A_S}\lesssim 350$~GeV. The limits from $A_S\rightarrow ZH, hA$ and $A \rightarrow h A_S$ are also stronger, comparing to $h_S\rightarrow ZA, HH, hH$ and $H \rightarrow h h_S, h_Sh_S$ in the left panel.

\section{Conclusions}
\label{sec:conclu}

The extension of the 2HDM models with a complex scalar singlet, motivated by NMSSM, Dark Matter consideration, and the 95 GeV excess,  offers extremely rich phenomenology in Higgs physics.   In this paper, we parameterize the 2HDM+S by the mass eigenvalues of the physical Higgses and the mixing angles, which provides a model-independent way to study the phenomenology. Our results can be applied to different 2HDM+S models with specific symmetry imposed on the Higgs potential by mapping the parameters into the mass eigenvalues and mixing angles.  We considered five benchmark scenarios in the Type-II Yukawa structure, with at most one mixing angle is set to be non-zero in each benchmark scenario.  We explored the 95\% C.L. constraints on the Higgs masses and mixing angles by considering the SM-like 125 GeV Higgs precision measurements, electroweak $STU$ constraints, direct search limits of the non-SM Higgses, and flavor constraints.

In the Higgs mass vs $\tan\beta$ space, we found that conventional channels of $\tau\tau$, followed by $bb$ in the large $\tan\beta$ region and $tt$ in the small $\tan\beta$ above the top threshold and $\gamma\gamma$ below the top threshold region provide the dominant constraints. However,  constraints on $m_{A/H/H^\pm}$ depend sensitively on the mass of the singlet $m_{h_S, A_S}$. While the reach for the conventional channels weaken when exotic channels such as $A/H \rightarrow Z h_S/ZA_S$ open, those exotic channels exclude large part of the parameter spaces, especially for moderate value of  $1<\tan\beta<7$   where the conventional channels can not contribute much.

We also studied the parameter space of Higgs mass vs mixing angle, and found that exotic channels of $A/H \rightarrow Z h_S/A_S$ as well as $H/h_S \rightarrow hh$ play a major role in constraining the range of the mixing angle.  For the $h_S-h_{125}$ mixing cases, $h_S \rightarrow VV$ could also be effective.

In the mixing angle vs $\tan\beta$ plane, while the conventional channels are more effective in constraining value of $\tan\beta$ across all values of the mixing angles, the exotic channels are more effective in constraining the large values of the mixing angles.  In the $m_{A,H,H^\pm}$ vs. $m_{h_S, A_S}$ plane, we found that the mass constraints for 2HDM-like Higgses mainly come from the conventional channel, while the mass constraints for the singlet Higgs mainly come from the exotic decays of $A$ and $H$, especially with large mixing angles.

In the $\tan\beta$ vs $\cos(\beta-\alpha)$ plane, we also studied the impact of the singlet mixing angles. The $h_S-h_{125}$ mixing angle has the largest impact on the constraints from the SM-like 125 GeV Higgs precision measurement, while $h_S-H$ mixing relax the constraints from the direct Higgs search channels.  

With the rich data available for both the Higgs precision measurements as well as direct BSM Higgs searches, our study set up a framework to systematically study the 2HDM+S parameter space given the current and future experimental searches.  While we only studied five benchmark scenarios with at most one mixing angle being nonzero, we identified the key features of the impact of a particular mixing angle, which helps to understand the  generic mixing cases in a more comprehensive study.

\section*{Acknowledgements}
CL, JL and WS are supported by the Natural Science Foundation of China (NSFC) under grant numbers 12305115, Shenzhen Science and Technology Program (Grant No. 202206193000001, 20220816094256002), Guangdong Provincial Key Laboratory of Gamma-Gamma Collider and Its Comprehensive Applications (2024KSYS001), and Guangdong Provincial Key Laboratory of Advanced Particle Detection Technology (2024B1212010005). JL is also supported by the Fundamental Research Funds for the Central Universities, and the Sun Yat-sen University Science Foundation. SS is supported by the Department of Energy under Grant No. DEFG02-13ER41976/DE-SC0009913.



\appendix
\section{Mass matrices}
For the most general 2HDM+S, the two doublet fields and one singlet field yield three neutral CP-even Higgses $H,h,h_S$,  two CP-odd Higgses $A,~A_S$, and one pair of charged Higgses $H^\pm$. After the electroweak symmetry breaking, the bilinear terms of the Higgs potential yield the mass matrices $M_S^2$ for CP-even states,  $M_P^2$ for the CP-odd states, and the charged Higgs mass.  The entries of the symmetric $M_S^2$ are given by
\begin{equation}
    \begin{split}
    M_{S11}^2 =& v^2(\lambda_1  \cos^2\beta+ \frac{3}{2}\lambda_6 \sin\beta\cos\beta -\frac{1}{2}\lambda_7\sin^2\beta\tan\beta) + m_{\phi}^2\sin^2\beta,\\
    M_{S22}^2 =& v^2(\lambda_2  \sin^2\beta -\frac{1}{2}\lambda_6\frac{\cos^2\beta}{\tan\beta}+\frac{3}{2}\lambda_7\sin\beta\cos\beta) + m_{\phi}^2\cos^2\beta, \\
    M_{S12}^2 =& ((\lambda_3+\lambda_4+\lambda_5+\frac{3}{2}[\lambda_6\cos^2\beta+\lambda_7\sin^2\beta])v^2 -m_{\phi}^2)\cos\beta\sin\beta,\\
    M_{S13}^2 =& v\left(2[\mu_{11}+(\lambda'_1+2\lambda'_4)v_S]  \cos\beta+(\mu_{12}+\mu_{21} +2[\lambda_3'+\lambda_6'+\lambda_7']v_S )\sin\beta\right),\\
    M_{S23}^2 =& v\left(2[\mu_{22}+(\lambda'_2+2\lambda'_5)v_S]  \sin\beta+(\mu_{12}+\mu_{21} +2[\lambda_3'+\lambda_6'+\lambda_7']v_S)\cos\beta\right),\\
    M_{S33}^2 =& \frac{2}{3}(\lambda''_1+4\lambda''_2+3\lambda''_3) v_S^2 + (\mu_{S1}+3\mu_{S2}) v_S \\&-\frac{v^2}{v_S}\left(\mu_{11}\cos^2\beta + (\mu_{12}+\mu_{21})\sin\beta\cos\beta + \mu_{22}\sin^2\beta \right).
    \end{split}
    \label{eq:masssmatrices}
\end{equation}
The entries of the symmetric $M_P^2$ are given by
\begin{equation}
\begin{split}
    M_{P11}^2 =& m_{\phi}^2-\left(\lambda_5+\frac{1}{2}(\frac{\lambda_6}{\tan\beta}+\lambda_7\tan\beta)\right) v^2,\\
    M_{P12}^2 =&( \mu_{21}-\mu_{12} + 2 [\lambda_7'-\lambda_6']v_S) v,\\
    M_{P22}^2 =& -4{m'_S}^2-\frac{4}{3}(\lambda''_1+\lambda''_2) v_S^2  -(3\mu_{S1}+\mu_{S2})v_S -2(\lambda_6' +\lambda_7')v^2\sin2\beta\\&-\frac{v^2}{v_S}\Big( (\mu_{11} + 4\lambda_4' v_S)\cos^2\beta + (\mu_{12}+\mu_{21})\cos\beta\sin\beta + (\mu_{22}+4\lambda_5' v_S)\sin^2\beta \Big).
\end{split}
\label{eq:masspmatrices}
\end{equation}
The charged Higgs mass is
\begin{equation}
        m_{H^\pm}^2 = \frac{1}{2}\left(\lambda_4 +\lambda_5 +\frac{\lambda_6}{\tan\beta}+\lambda_7\tan\beta\right)v^2 -m_{\phi}^2.
        \label{eq:masscmatrices}
\end{equation}
The $m_{\phi}^2$ parameter is defined by
\begin{equation}
    m_{\phi}^2 = \frac{m_{12}^2-(\mu_{12}+\mu_{21}+(\lambda_3'+\lambda_6'+\lambda_7')v_S) v_S}{\sin\beta\cos\beta}.\label{eq:mphi}
\end{equation}

The mass eigenvalues can be obtained by the diagonalization of the mass matrices with the mixing matrix $R$:
\begin{equation}
    M_{Sij}^2 = \sum_k m_{h_k}^2R_{ki}R_{kj}.
\end{equation}
Therefore, we can use the Higgs boson masses and mixing angles to solve the parameters in the interaction basis via the mass matrix entries: 
\begin{equation}
\begin{split}
    &\lambda_1 =\frac{1}{v^2}\left( \frac{M_{S11}^2}{\cos^2\beta}-m_{\phi}^2\tan^2\beta\right)-\frac{3}{2}\lambda_6\tan\beta + \frac{\lambda_7}{2}\tan^3\beta,\\
    &\lambda_2 =\frac{1}{v^2}\left( \frac{M_{S22}^2}{\sin^2\beta}-\frac{m_{\phi}^2}{\tan^2\beta}\right)+\frac{\lambda_6}{2\tan^3\beta}-\frac{3\lambda_7}{2\tan\beta},\\
    &\lambda_3+\lambda_4+\lambda_5 = \frac{1}{v^2}\left( m_{\phi}^2 +\frac{M_{S12}^2}{\sin\beta\cos\beta}\right)-\frac{3}{2}(\lambda_6/\tan\beta+\lambda_7\tan\beta),\\
    &\lambda_5 = \frac{m_{\phi}^2-M_{P11}^2}{v^2}-\frac{1}{2}(\lambda_6/\tan\beta + \lambda_7\tan\beta),\\
        &\lambda_4+\lambda_5 =2 \frac{m_{H^\pm}^2-m_{\phi}^2}{v^2}-(\lambda_6/\tan\beta + \lambda_7\tan\beta),\\
    &\lambda'_1+2\lambda'_4 = \frac{M_{S13}^2 - v[2\mu_{11}\cos\beta + (\mu_{12}+\mu_{21}+2[\lambda_3'+\lambda_6'+\lambda_7']v_S) \sin\beta]}{2v v_S \cos\beta},\\
    &\lambda'_2+2\lambda'_5 = \frac{M_{S23}^2 - v[2\mu_{22}\sin\beta + (\mu_{12}+\mu_{21}+2[\lambda_3'+\lambda_6'+\lambda_7']v_S) \cos\beta]}{2v v_S \sin\beta},\\
    &\frac{\lambda''_1+4\lambda''_2 + 3\lambda''_3}{3} = \frac{1}{2 v_S^2}\Big( M_{S33}^2 - (\mu_{S1}+3\mu_{S2})v_S \\&~~~~+ \frac{v^2}{2v_S} (\mu_{11}\cos^2\beta + {(\mu_{12}+\mu_{21})}\sin\beta\cos\beta +\mu_{22}\sin^2\beta) \Big),\\
    &\lambda_7'-\lambda_6' = \frac{1}{2v_S}\left( \frac{M_{P12}^2}{v}+\mu_{12}-\mu_{21} \right),\\
    &\frac{4}{3}(\lambda''_1+\lambda''_2) =-M_{P22}^2 -4\frac{{m'_S}^2}{v_S^2}   -\frac{3\mu_{S1}+\mu_{S2}}{v_S} -2(\lambda_6' +\lambda_7')\frac{v^2}{v_S^2}\sin2\beta\\&-\frac{v^2}{v_S^3}\Big( (\mu_{11} + 4\lambda_4' v_S)\cos^2\beta + (\mu_{12}+\mu_{21})\cos\beta\sin\beta + (\mu_{22}+4\lambda_5' v_S)\sin^2\beta \Big).
\end{split}\label{eq:basischange}
\end{equation}


\section{Higgs boson couplings}
\label{sec:hcpls}

In Table~\ref{tab:hff_hvcoups}, we  present the reduced Higgs boson couplings to the fermions and gauge bosons.
\begin{table}[h]
    \centering
    \resizebox{\linewidth}{!}{
    \begin{tabular}{lcccccc}
    \hline
    & Couplings& Case-0&   Case-I& Case-II&    Case-III& Case-IV\\
    \hline
$c_{HVV}$& $c_{\beta-\alpha}c_{\alpha_{HS}} $&    0&    $c_{\beta-\alpha}$&   0&  0& 0   \\
\hline
$c_{Huu}$& $c_{HVV}-s_{\beta-\alpha}c_{\alpha_{HS}}/t_\beta$ &    -$1/t_\beta$&    $c_{\beta-\alpha}-s_{\beta-\alpha}/t_\beta$&   -$1/t_\beta$&  -$c_{\alpha_{HS}}/t_\beta$& -$1/t_\beta$   \\
$c_{Hdd}$ & $c_{HVV}+s_{\beta-\alpha}c_{\alpha_{HS}}t_\beta$ &    $t_\beta$&    $c_{\beta-\alpha}+s_{\beta-\alpha}t_\beta$&  $t_\beta$&  $c_{\alpha_{HS}}t_\beta$& $t_\beta$   \\
$c_{Hll}$& $c_{HVV}+s_{\beta-\alpha}c_{\alpha_{HS}}t_\beta$ &    $t_\beta$&    $c_{\beta-\alpha}+s_{\beta-\alpha}t_\beta$&  $t_\beta$&  $c_{\alpha_{HS}}t_\beta$& $t_\beta$   \\
\hline
$c_{hVV}$& $s_{\beta-{\alpha}}c_{\alpha_{hS}} - c_{\beta-{\alpha}}s_{\alpha_{HS}}s_{\alpha_{hS}} $&    1&    $s_{\beta-\alpha}$&   $c_{\alpha_{hS}}$&   1& 1   \\
\hline
$c_{huu}$& $c_{hVV}+(c_{\beta-\alpha}c_{\alpha_{hS}}+s_{\beta-\alpha}s_{\alpha_{hS}}s_{\alpha_{HS}})/t_\beta$ &    1&    $s_{\beta-\alpha}+c_{\beta-\alpha}/t_\beta$&   $c_{\alpha_{hS}}$&  1& 1   \\
$c_{hdd}$& $c_{hVV}-(c_{\beta-\alpha}c_{\alpha_{hS}}+s_{\beta-\alpha}s_{\alpha_{hS}}s_{\alpha_{HS}})t_\beta$ &    1&    $s_{\beta-\alpha}-c_{\beta-\alpha}t_\beta$&   $c_{\alpha_{hS}}$&  1& 1   \\
$c_{hll}$& $c_{hVV}-(c_{\beta-\alpha}c_{\alpha_{hS}}+s_{\beta-\alpha}s_{\alpha_{hS}}s_{\alpha_{HS}})t_\beta$ &    1&    $s_{\beta-\alpha}-c_{\beta-\alpha}t_\beta$&   $c_{\alpha_{hS}}$&  1& 1   \\
\hline
$c_{h_SVV}$& $-s_{\beta-\alpha}s_{\alpha_{hS}} - c_{\beta-\alpha}s_{\alpha_{HS}}c_{\alpha_{hS}}$&    0&    0&   $-s_{\alpha_{hS}}$&   0& 0   \\
\hline
$c_{h_Suu}$& $c_{h_SVV}-(c_{\beta-\alpha}s_{\alpha_{hS}}-s_{\beta-\alpha}c_{\alpha_{hS}}s_{\alpha_{HS}})/t_\beta$&    0&    0&   $-s_{\alpha_{hS}}$&   $s_{\alpha_{HS}}/t_\beta$& 0   \\
$c_{h_Sdd}$& $c_{h_SVV}+(c_{\beta-\alpha}s_{\alpha_{hS}}-s_{\beta-\alpha}c_{\alpha_{hS}}s_{\alpha_{HS}})t_\beta$&    0&    0&   $-s_{\alpha_{hS}}$&   $-s_{\alpha_{HS}}t_\beta$& 0   \\
$c_{h_Sll}$& $c_{h_SVV}+(c_{\beta-\alpha}s_{\alpha_{hS}}-s_{\beta-\alpha}c_{\alpha_{hS}}s_{\alpha_{HS}})t_\beta$&    0&    0&   $-s_{\alpha_{hS}}$&   $-s_{\alpha_{HS}}t_\beta$& 0   \\
\hline
$c_{Auu}$& ${c_{{\alpha_{AS}}}}{/t_\beta}$&    $1/t_\beta$&    $1/t_\beta$&   $1/t_\beta$&   $1/t_\beta$& ${c_{{\alpha_{AS}}}}{/t_\beta}$   \\
$c_{Add}$& $-{c_{\alpha_{AS}}}t_\beta $&    $-t_\beta$&    $-t_\beta$&   $-t_\beta$&   $-t_\beta$& $-{c_{\alpha_{AS}}}t_\beta $   \\
$c_{All}$& $-{c_{\alpha_{AS}}}t_\beta $&    $-t_\beta$&    $-t_\beta$&   $-t_\beta$&   $-t_\beta$& $-{c_{\alpha_{AS}}}t_\beta $   \\
\hline
$c_{a_Suu}$& $-{s_{{\alpha_{AS}}}}{/t_\beta}$&    0&    0&   0&   0& $-{s_{{\alpha_{AS}}}}{/t_\beta}$   \\
$c_{a_Sdd}$& ${s_{\alpha_{AS}}}t_\beta$&    0&    0&   0&   0& ${s_{\alpha_{AS}}}t_\beta$   \\
$c_{a_Sll}$& ${s_{\alpha_{AS}}}t_\beta$&    0&    0&   0&   0& ${s_{\alpha_{AS}}}t_\beta$   \\
\hline
$c_{A H Z}$& $ -c_{\alpha_{AS}}c_{\alpha_{HS}}s_{\beta-{\alpha}}$& -1& $-s_{\beta-{\alpha}}$& -1& $-c_{\alpha_{HS}}$& $-c_{\alpha_{AS}}$\\
$c_{AhZ}$& $ c_{\alpha_{AS}}\Big(c_{\beta-{\alpha}}c_{\alpha_{hS}} + s_{\beta-{\alpha}}s_{\alpha_{HS}}s_{\alpha_{hS}} \Big)$& 0&  $c_{\beta-{\alpha}}$& 0& 0& 0\\
$c_{Ah_S Z}$& $ -c_{\alpha_{AS}}\Big(c_{\beta-{\alpha}}s_{\alpha_{hS}} - s_{\beta-{\alpha}}s_{\alpha_{HS}}c_{\alpha_{hS}} \Big)$& 0&  0& 0& $s_{\alpha_{HS}}$& 0\\
$c_{A_S HZ}$& $s_{\alpha_{AS}}c_{\alpha_{HS}}s_{\beta-{\alpha}}$& 0&  0& 0& 0& $s_{\alpha_{AS}}$\\
$c_{A_S h Z}$& $ -s_{\alpha_{AS}}\Big(c_{\beta-{\alpha}}c_{\alpha_{hS}} + s_{\beta-{\alpha}}s_{\alpha_{HS}}s_{\alpha_{hS}} \Big)$& 0&  0& 0& 0&  0\\
$c_{A_S h_S Z}$& $ s_{\alpha_{AS}}\Big(c_{\beta-{\alpha}}s_{\alpha_{hS}} - s_{\beta-{\alpha}}s_{\alpha_{HS}}c_{\alpha_{hS}} \Big)$&0&  0& 0& 0& 0\\
\hline
$c_{H H^\pm W^\mp}$& $-i c_{\alpha_{HS}}s_{\beta-{\alpha}}$& -i& $-is_{\beta-{\alpha}}$& -i& $-ic_{\alpha_{HS}}$& -i\\
$c_{h H^\pm W^\mp}$& $i\Big(c_{\beta-{\alpha}} c_{\alpha_{hS}} + s_{\beta-{\alpha}}s_{\alpha_{HS}}s_{\alpha_{hS}} \Big)$& 0& $ic_{\beta-{\alpha}}$& 0& 0& 0\\
$c_{h_S H^\pm W^\mp}$& $-i\Big(c_{\beta-{\alpha}} s_{\alpha_{hS}} - s_{\beta-{\alpha}}s_{\alpha_{HS}}c_{\alpha_{hS}} \Big)$& 0& 0& 0& $-is_{\alpha_{HS}}$& 0\\
$c_{A H^\pm W^\mp}$& $c_{\alpha_{AS}}$& 1& 1& 1& 1& $c_{\alpha_{AS}}$\\
$c_{A_S H^\pm W^\mp}$& $-s_{\alpha_{AS}}$& 0& 0& 0& 0& $-s_{\alpha_{AS}}$\\
\hline
\end{tabular}}
\caption{The reduced Higgs couplings to fermions and gauge bosons in the general form and under five benchmark cases.}
\label{tab:hff_hvcoups}
\end{table}

The trilinear Higgs couplings can be obtained by the third derivative of the Higgs potential
\begin{equation}
    ig_{\phi_i\phi_j\phi_k}=\frac{\partial^3V}{\partial\phi_i\partial\phi_j\partial\phi_k}\Big|_{\Phi_i = v_i}
\end{equation}
By applying the basis change in Eq.~\eqref{eq:basischange}, the trilinear Higgs couplings can be expressed in terms of the masses and mixings as follows
\begin{equation}
    \begin{split}
        g_{h_i h_j h_k} =& -\frac{m_{h_i}^2 + m_{h_j}^2 + m_{h_k}^2}{v}\Big( \frac{R_{i1}R_{j1}R_{k1}}{c_\beta} + \frac{R_{i2}R_{j2}R_{k2}}{s_\beta} + R_{i3}R_{j3}R_{k3}\frac{v}{v_S} \Big) \\
        & + \frac{m_{\phi}^2}{2v}\left(\Lambda^{111}_{ijk}\frac{ s_\beta^2}{c_\beta } + \Lambda^{222}_{ijk}\frac{ c_\beta^2}{s_\beta } - {\Lambda_{ijk}^{122} c_\beta  - \Lambda_{ijk}^{112}}s_\beta \right) \\
        &  + (\mu_{12}+\mu_{21} + 2[\lambda_3'+\lambda_6'+\lambda_7']v_S) \Big(\frac{\Lambda_{113}t_\beta + \Lambda_{223}/t_\beta}{2} -\Lambda_{ijk}^{123}\Big)\\
        &  + \left(\mu_{11}+\mu_{22}+\frac{\mu_{12}+\mu_{21}}{2}\right)\frac{v}{v_S}\frac{s_\beta \Lambda_{ijk}^{133} + c_\beta \Lambda_{ijk}^{233}}{2} \\
        &  - 3\frac{v^2}{v_S^2}\Big(\mu_{11}c_\beta^2 + (\mu_{12}+\mu_{21})s_\beta c_\beta + \mu_{22}s_\beta^2\Big) R_{i3}R_{j3}R_{k3}   + (\mu_{S1}+3\mu_{S2})R_{i3}R_{j3}R_{k3}\\
        & -\frac{3}{2}v(\lambda_6 c^2_\beta - \lambda_7 s^2_\beta)s_\beta c_\beta\left( \frac{R_{i2}}{s_\beta}-  \frac{R_{i1}}{c_\beta}\right)\left( \frac{R_{j2}}{s_\beta}-  \frac{R_{j1}}{c_\beta}\right)\left( \frac{R_{k2}}{s_\beta}-  \frac{R_{k1}}{c_\beta}\right),
    \end{split}
    \label{eq:hhh}
\end{equation}
where
\begin{equation}
    \Lambda_{ijk}^{abc}=R_{ia}R_{jb}R_{kc}+ R_{ib}R_{jc}R_{ka} + R_{ic}R_{ja}R_{kb} + R_{ia}R_{jc}R_{kb}+R_{ib} R_{ja}R_{kc} + R_{ic}R_{jb}R_{ka}.
\end{equation}
In Table~\ref{tab:hhh_01234}, we present the formula of trilinear CP-even Higgs couplings in five benchmark cases.
The general formula of $g_{h_ia_ja_k}$ couplings is given by
\begin{equation}
    \begin{split}
        g_{h_i a_j a_k}=& \Bigg[ \frac{m_{\phi}^2 }{v}  \left( \frac{R_{i2}}{s_\beta}+\frac{R_{i1}}{c_\beta}\right)-\frac{m_{h_i}^2}{v}  \left( \frac{R_{i1}s^2_\beta}{c_\beta}+\frac{R_{i2}c^2_\beta}{s_\beta}\right)  -\frac{m_{a_j}^2+m_{a_k}^2}{v}(R_{i1}c_\beta + 
        R_{i2}s_\beta)
        \\& +\frac{v}{2}(\lambda_7 t_\beta -\lambda_6/t_\beta)\left(\frac{R_{i2}}{s_\beta}-\frac{R_{i1}}{c_\beta}\right) + \frac{\mu_{12}+\mu_{21}+2(\lambda_3'+\lambda_6'+\lambda_7') v_S}{ s_\beta c_\beta} R_{i3} \Bigg]R^A_{j1}R^A_{k1}\\
       &  + \Bigg[- \Bigg(\frac{m_{h_i}^2 + m_{a_j}^2+m_{a_k}^2}{v_S}   +8\frac{{m_S'}^2}{v_S} +\frac{v^2}{v_S^2}\Big((3\mu_{11}+8\lambda_4')c_\beta^2 +(3\mu_{22}+8\lambda_5')s^2_\beta\Big)\\&+ 3\mu_{S1}+\mu_{S2}\Bigg) R_{i3}   +(\mu_{12}+\mu_{21})\left (\frac{v}{v_S}(R_{i1 }s_\beta + R_{i2} c_\beta) -3 \frac{v^2}{v_S^2}s_\beta c_\beta R_{i3} \right)
     \\&+2\frac{v}{v_S} \Big((\mu_{11}+4\lambda_4' v_S)c_\beta R_{i1} + (\mu_{22}+4\lambda_5' v_S)s_\beta R_{i2} \Big)
      \\&+4(\lambda_6'+\lambda_7') v \left( R_{i1} s_\beta + R_{i2} c_\beta - \frac{v}{v_S}R_{i3}s_{2\beta} \right)   \Bigg]R^A_{j2} R^A_{k2}\\
       &+(\mu_{21}-\mu_{12})\frac{v}{v_S}R_{i3}(R^A_{j1}R^A_{k2}+R^A_{j1}R^A_{k2}).
        \label{eq:haa}
    \end{split}
\end{equation}
The general formula of the charged Higgs bosons couplings $g_{h_iH^+H^-}$ is given by
\begin{equation}
    \begin{split}
         g_{h_i H^+ H^-} =& \frac{m_\phi^2}{v}\left( \frac{R_{i2}}{s_\beta}+\frac{R_{i1}}{c_\beta} \right)-\frac{m_{h_i}^2}{v}\left( R_{i1} \frac{s_\beta^2}{c_\beta}+R_{i2}\frac{c_\beta^2}{s_\beta} \right)-\frac{2m_{H^\pm}^2}{v}(R_{i1}c_\beta + R_{i2}s_\beta)\\&  + \frac{{\mu_{12}+\mu_{21}}+2(\lambda_3'+\lambda_6'+\lambda_7')v_S}{s_\beta c_\beta}R_{i3}+\frac{v}{2}(\lambda_7 t_\beta -\lambda_6/t_\beta)\left(\frac{R_{i2}}{s_\beta}-\frac{R_{i1}}{c_\beta}\right).
    \end{split}\label{eq:hhphm}
\end{equation}

\begin{table}[h]
    \centering
    \resizebox{\linewidth}{!}{\begin{tabular}{llllll}
        \hline
    & 0& I& II& III& IV\\
    \hline
$ g_{h h h}$ & $-\frac{3 m_{h}^2}{v}$& $g_{hhh}^0 s_{\beta-\alpha}$& $g_{hhh}^0 c^3_{\alpha_{h_S}} - \frac{(9m_h^2 +m_{A_S}^2)s^3_{\alpha_{hS}}}{3v_S}$&   $g_{hhh}^0$&    $g_{hhh}^0$\\
& & $-6\frac{m_h^2-m_\phi^2 }{v}c^3_{\beta-\alpha}(t_{\beta-\alpha}+\frac{1}{t_{2\beta}})$  & &&\\\\

$g_{H h h }$& 0& $-c_{\beta-\alpha}[\frac{m_\phi^2}{v} $&   0&  0&  0\\&&$+\frac{m_H^2+2m_h^2 - 3m_\phi^2}{v}c_{2(\beta-\alpha)}(1-\frac{t_{2(\beta-\alpha)}}{t_{2\beta}})  ]$&&&\\\\  

$g_{hHH}$& $- \frac{2m_H^2+m_h^2-2m_\phi^2}{v}$&  $g_{hHH}^0 s^3_{\beta-\alpha}(1+\frac{1}{t_{2\beta}t_{\beta-\alpha}}) $&    $g_{hHH}^0 c_{\alpha_{hS}}$&  $g_{hHH}^0c_{\alpha_{HS}}$&   $g_{hHH}^0 + \frac{2(m_A^2-m_{A_S}^2)s^2_{\alpha_{AS}}}{v}$\\&&$- \frac{m_\phi^2}{v} s_{\beta-\alpha}c^2_{\beta-\alpha}(1-\frac{1}{t_{2\beta}t_{\beta-\alpha}})$&&&
 \\\\

$g_{HHH}$& $6\frac{m_H^2-m_\phi^2}{v t_{2\beta}}$&  $g_{HHH}^0 s^3_{\beta-\alpha}$ &  $g_{HHH}^0$&   $g_{HHH}^0c^3_{\alpha_{HS}}-\frac{9m_H^2+m_{A_S}^2}{3 v_S}s^3_{\alpha_{HS}}$& $g_{HHH}^0+ \frac{6(m_A^2-m_{A_S}^2)s^2_{\alpha_{AS}}}{vt_{2\beta}}$\\&&$+3c_{\beta-\alpha}\frac{2(m_\phi^2-m_H^2) s_{\beta-\alpha}^2 -m_H^2}{v}$&&&\\\\

$g_{h_S h_S h_S}$&  $-\frac{9m_{h_S}^2+m_{A_S}^2}{3v_S}$& $g_{h_sh_Sh_S}^0$&     $g_{h_S h_S h_S}^0c^3_{\alpha_{hS}}$& $g_{h_S h_S h_S}^0c^3_{\alpha_{HS}}-6\frac{m_{h_S}^2-m_A^2}{v t_{2\beta}}s^3_{\alpha_{HS}}$&   $g_{h_Sh_Sh_S}^0-\frac{m_A^2-m_{A_S}^2}{3v_S}s^2_{\alpha_{AS}}$\\
&&&$+\frac{3 m_{h_S}^2 s^3_{\alpha_{h_S}}}{v}$&& $-5v\frac{m_A^2+m_{A_S}^2}{6v_S^2}s_{2\beta}s_{2\alpha_{AS}}$\\\\

$g_{h h_S h_S}$&  0&0&   $\Big[- \frac{m_{h}^2+2m_{h_S}^2}{2}(\frac{c_{\alpha_{hS}}}{v_S}+\frac{s_{\alpha_{hS}}}{v}) $ & $-\frac{m_{h}^2+2m_{h_S}^2-2m_A^2}{v} s^2_{\alpha_{HS}}$&   $\frac{m_{A_S}^2-m_A^2}{2v_S}s_{2\beta}s_{2\alpha_{AS}}$   \\
&&& $-\frac{m_{A_S}^2 c_{\alpha_{hS}}}{6 v_S}\Big]s_{2\alpha_{hS}}$&&\\\\

$g_{h h h_S}$&  0&0&   $\Big[ \frac{2m_{h}^2+m_{h_S}^2}{2}(\frac{c_{\alpha_{hS}}}{v}-\frac{s_{\alpha_{hS}}}{v_S})$& 0&  0\\
&& &  $-\frac{m_{A_S}^2 s_{\alpha_{hS}}}{6 v_S} \Big]s_{2\alpha_{hS}}$&\\\\

$g_{H h_S h_S}$&  0&0&   0& $\Big((m_H^2+2m_{h_S}^2)[\frac{s_{\alpha_{HS}}}{vt_{2\beta}}-\frac{c_{\alpha_{HS}}}{2v_S}]$&    $\frac{m_A^2-m_{A_S}^2}{2 v_S}c_{2\beta}s_{2\alpha_{AS}}$\\
&&&& $-[\frac{3m_{A}^2s_{\alpha_{HS}}}{vt_{2\beta} } +\frac{m_{A_S}^2c_{\alpha_{HS}}}{6v_S}]\Big) s_{2\alpha_{HS}}$
\\\\

$g_{H H h_S}$&  0&0&   $\frac{m_{h_S}^2+2m_{H}^2-2m_{A}^2}{v}s_{\alpha_{hS}}$& $\Big([\frac{3m_{A}^2c_{\alpha_{HS}}}{vt_{2\beta}}-\frac{m_{A_S}^2 s_{\alpha_{HS}}}{6v_S}]$& $\frac{m_{A_S}^2-m_A^2}{4v}s_{2\beta}s_{2\alpha_{AS}}$\\
&&&& $-(2m_H^2+m_{h_S}^2)[\frac{c_{\alpha_{HS}}}{vt_{2\beta}}+\frac{s_{\alpha_{HS}}}{2v_S}]\Big) s_{2\alpha_{HS}}$\\\\

$g_{H h h_S}$&  0&0&   0& $\frac{m_{h_S}^2+m_H^2+m_h^2-2m_A^2}{v}s_{\alpha_{HS}}c_{\alpha_{HS}}$& 0\\\\

\hline
    \end{tabular}}
    \caption{The trilinear Higgs coupling $g_{h_i h_j h_k}$ in five fundamental cases.}
    \label{tab:hhh_01234}
\end{table}

In Table~\ref{tab:haamp}, we present the $g_{h_i a_j a_k}$ and $g_{h_i H^+H^-}$ in five benchmark cases. 
\begin{table}[h]
    \centering
    \resizebox{\textwidth}{!}{ \begin{tabular}{lllllll}
    \hline
    &   0&  I&  II& III&    IV\\
    \hline
$g_{h H^+H^-}$& $-\frac{m_h^2+2m_{H^\pm}^2-2m_\phi^2}{v}$&  $g_{h H^+ H^-}^0 s_{\beta-\alpha} + 2\frac{m_\phi^2 -m_h^2}{vt_{2\beta}} c_{\beta-\alpha}$&   $g^0_{hH^+H^-} c_{\alpha_{hS}}$&  $g^0_{hH^+H^-}$&  $g^0_{hH^+H^-}$\\\\
$g_{H H^+H^-}$& $2\frac{m_H^2-m_\phi^2}{vt_{2\beta}}$& $g_{H H^+ H^-}^0 s_{\beta-\alpha} - \frac{m_H^2+2m_{H^\pm}^2-2m_\phi^2}{v}c_{\beta-\alpha}$&   $g^0_{HH^+H^-} $&  $g^0_{HH^+H^-} c_{\alpha_{HS}}$& $g^0_{HH^+H^-}$ \\\\
$g_{h_S H^+H^-}$& 0&0&   $\frac{m_{h_S}^2+2m_{H^\pm}^2-2m_\phi^2}{v}s_{\alpha_{hS}}$& $-2\frac{m_{h_S}^2-m_\phi^2}{vt_{2\beta}}s_{\alpha_{HS}}$&  $-\frac{m_A^2-m_{A_S}^2}{vs_{2\beta}}s_{2\alpha_{AS}}$ \\\\

$g_{HAA}$&  $2\frac{m_H^2-m_\phi^2}{vt_{2\beta}}$&     $g_{HAA}^0 s_{\beta-\alpha} + \frac{2m_\phi^2-2m_A^2-m_H^2}{v}c_{\beta-\alpha} $&  $g_{HAA}^0$&  $g_{HAA}^0 c_{\alpha_{HS}}$& $\frac{m_A^2-m_{A_S}^2}{2}s_{2\alpha_{AS}}(\frac{s_{2\alpha_{AS}}}{vt_{2\beta}} + \frac{c_{2\beta}s^2_{\alpha_{AS}}}{v_S})$\\
&&&&& $+g_{HAA}^0c^2_{\alpha_{AS}}$\\\\

$g_{HA_SA_S}$&  0&0&   $0$& $-\frac{m_{A_S}^2+m_H^2}{v_S}s_{\alpha_{HS}}$& $(m_{A}^2 - m_{A_S}^2)(\frac{2s^4_{\alpha_{AS}}}{vt_{2\beta}}+\frac{c_{2\beta}s_{\alpha_{AS}}c^3_{\alpha_{AS}}}{v_S}) $\\
&&&&& $+g_{HAA}^0 s^2_{\alpha_{AS}} $\\\\
$g_{H A A_S }$& 0&0&   0& 0& $ (m_A^2-m_{A_S}^2)s_{2\alpha_{AS}}( \frac{c_{2\beta} s_{2\alpha_{AS}}}{4v_S}-\frac{s^2_{\alpha_{AS}}}{vt_{2\beta}})$\\
&&&&& $-\frac{s_{2\alpha_{AS}}}{4}g^0_{HAA}$\\\\

$g_{hAA}$&  $-\frac{2m_A^2+m_h^2-2m_\phi^2}{v}$&   $g_{hAA}^0 s_{\beta-\alpha} + 2\frac{m_\phi^2 -m_h^2}{vt_{2\beta}} c_{\beta-\alpha}$&  $g_{hAA}^0c_{\alpha_{hS}}$&  $g_{hAA}^0$& $\frac{m_{A_S}^2-m_A^2}{2}s_{2\alpha_{AS}}(\frac{s_{2\alpha_{AS}}}{v}+\frac{s_{2\beta}s^2_{\alpha_{AS}}}{v_S})$\\
&&&&& $+g_{hAA}^0 c^2_{\alpha_{AS}}$\\\\

$g_{hA_SA_S}$&  0&0&   $-\frac{m_{A_S}^2+m_h^2}{v_S}s_{\alpha_{hS}}$& 0& $\frac{m_{A}^2-m_{A_S}^2}{2}(\frac{s^2_{2\alpha_{AS}}}{v}-\frac{s_{2\beta}s_{2\alpha_{AS}}c^2_{\alpha_{AS}}}{v_S})$\\
&&&&&$+g_{hAA}^0 s^2_{\alpha_{AS}}$\\\\
$g_{h A A_S }$& 0&0&   0& 0& $\frac{m_{A_S}^2-m_A^2}{4}(\frac{s_{4\alpha_{AS}}}{v}+\frac{s^2_{2\alpha_{AS}}s_{2\beta}}{v_S})$\\&&&&&$-\frac{s_{2\alpha_{AS}}}{2}g_{hAA}^0$\\\\

$g_{h_SAA}$&    0&0&   $\frac{2m_A^2+m_{h_S}^2-2m_\phi^2}{v}s_{\alpha_{hS}}$& $-2\frac{m_{h_S}^2-m_\phi^2}{vt_{2\beta}}s_{\alpha_{HS}}$&  $\frac{m_A^2-m_{A_S}^2}{v}(\frac{v^2 s_{2\beta} s^3_{\alpha_{AS}}c_{\alpha_{AS}}}{v_S^2}-\frac{vs^4_{\alpha_{AS}}}{v_S}-\frac{s_{2\alpha_{AS}}c^2_{\alpha_{AS}}}{s_{2\beta}})$\\&&&&&$+g_{h_SA_SA_S}^0s^2_{\alpha_{AS}}$\\\\
$g_{h_SA_S A_S}$&  $-\frac{m_{A_S}^2+m_{h_S}^2}{v_S}$& $g_{h_SA_SA_S}^0$& $g_{h_SA_SA_S}^0$& $g_{h_SA_SA_S}^0c_{\alpha_{HS}}$&  $\frac{
m_A^2-m_{A_S}^2
}{v}s_{2\alpha_{AS}}(\frac{v
^2 s_{2\beta}c^2_{\alpha_{AS}}}{2v_S^2}-\frac{vs_{2\alpha_{AS}}}{4v_S}-\frac{s^2_{\alpha_{AS}}}{s_{2\beta}})$\\&&&&& $+g_{h_SA_SA_S}^0c^2_{\alpha_{AS}}$\\\\
$g_{h_SAA_S}$&  0&  0&  0& 0& $-\frac{m_A^2-m_{A_S}^2}{2v}s_{2\alpha_{AS}}(\frac{s_{\alpha_{AS}}}{s_{2\beta}}+\frac{v \,s^2_{\alpha_{AS}}}{v_S}+\frac{v^2 s_{2\alpha_{AS}}s_{2\beta}}{2v_S^2})$\\&&&&& $+g^0s_{\alpha_{AS}}c_{\alpha_{AS}}$\\
\hline
    \end{tabular}}
    \caption{The Higgs couplings $g_{h_i H^+H^-}$ and $g_{h_ia_ja_k}$ in five fundamental cases.}
    \label{tab:haamp}
\end{table}

\newpage
\clearpage

\bibliographystyle{JHEP}
\bibliography{ref}

\end{document}